\documentclass[preprintnumbers, floatfix, letterpaper, aps, prd, epsfig,nofootinbib,
twocolumn
]{revtex4-1}
\usepackage{bm,graphicx,dcolumn,epstopdf,epsf, latexsym,mathbbol, amssymb,amsmath,color,slashed, mathrsfs,mathcomp,simplewick}
\usepackage[center]{subfigure}
\usepackage{multirow}
\usepackage{makecell}
\usepackage[colorlinks,linkcolor=blue,citecolor=blue,urlcolor=blue]{hyperref}
\usepackage{CJK}
\usepackage{gensymb}

\begin{document}
\allowdisplaybreaks
 \newcommand{\bq}{\begin{equation}}
 \newcommand{\eq}{\end{equation}}
 \newcommand{\bqn}{\begin{eqnarray}}
 \newcommand{\eqn}{\end{eqnarray}}
 \newcommand{\nb}{\nonumber}
 \newcommand{\lb}{\label}
 \newcommand{\f}{\frac}
 \newcommand{\p}{\partial}
\newcommand{\PRL}{Phys. Rev. Lett.}
\newcommand{\PLB}{Phys. Lett. B}
\newcommand{\PRD}{Phys. Rev. D}
\newcommand{\CQG}{Class. Quantum Grav.}
\newcommand{\JCAP}{J. Cosmol. Astropart. Phys.}
\newcommand{\JHEP}{J. High. Energy. Phys.}
\newcommand{\red}{\textcolor{red}}
\newcommand{\blue}{\textcolor{blue}}



\title{Constraining parity and Lorentz violations in gravity with future ground- and space-based gravitational wave detectors}

\author{Bo-Yang Zhang${}^{a, b, c}$}
\email{zhangby@stumail.neu.edu.cn}

\author{Tao Zhu${}^{b, c}$}
\email{Corresponding author: zhut05@zjut.edu.cn}

\author{Jian-Ming Yan${}^{b, c}$}
\email{yanjm@zjut.edu.cn}

\author{Jing-Fei Zhang${}^{a}$}
\email{jfzhang@mail.neu.edu.cn}

\author{Xin Zhang${}^{a,d,e}$}
\email{Corresponding author: zhangxin@mail.neu.edu.cn}

\affiliation{
${}^{a}$ Key Laboratory of Cosmology and Astrophysics (Liaoning) \& College of Sciences, Northeastern University, Shenyang 110819, China\\
${}^{b}$ Institute for Theoretical Physics and Cosmology, Zhejiang University of Technology, Hangzhou, 310032, China\\
${}^{c}$ United Center for Gravitational Wave Physics (UCGWP), Zhejiang University of Technology, Hangzhou, 310032, China\\
${}^{d}$ Key Laboratory of Data Analytics and Optimization for Smart Industry (Ministry of Education), Northeastern University, Shenyang 110819, China\\
${}^{e}$ National Frontiers Science Center for Industrial Intelligence and Systems Optimization, Northeastern University, Shenyang 110819, China}

\begin{abstract}

The future ground- and space-based gravitational wave (GW) detectors offer unprecedented opportunities to test general relativity (GR) with greater precision. In this work, we investigate the capability of future ground-based GW detectors, the Einstein Telescope (ET) and the Cosmic Explorer (CE), and space-based GW detectors, LISA, Taiji, and TianQin, for constraining parity and Lorentz violations in gravity. We inject several typical GW signals from compact binary systems into GW detectors and perform Bayesian inferences with the modified waveforms with parity and Lorentz-violating effects. These effects are modeled in the amplitude and phase corrections to the GW waveforms with their frequency-dependence described by factors $\beta_{\nu}$, $\beta_{\mu}$, $\beta_{\bar \nu}$, and $\beta_{\bar \mu}$. Our results show that the combined observations of ET and CE will impose significantly tighter bounds on the energy scale of parity and Lorentz violations ($M_{\rm PV}$ and $M_{\rm LV}$) compared to those given by LIGO-Virgo-KAGRA (LVK) detectors. For cases with positive values of $\beta_{\nu}$, $\beta_{\mu}$, $\beta_{\bar \nu}$, and $\beta_{\bar \mu}$, the constraints on $M_{\rm PV}$ and $M_{\rm LV}$ from ground-based detectors are tighter than those from the space-based detectors. For the $\beta_{\mu} = -1$ case, space-based GW detectors provide constraints on $M_{\rm PV}$ that are better than current LVK observations and comparable to those from ET and CE. Additionally, space-based detectors exhibit superior sensitivity in constraining $M_{\rm LV}$ for $\beta_{\bar \mu} = -2$ case, which is approximately three orders of magnitude tighter than those from ground-based GW detectors. This scenario also enables bounds on the graviton mass at $m_g \lesssim 10^{-35}\; {\rm GeV}$. These findings highlight the promising role of future GW observatories in probing fundamental physics beyond GR.

\end{abstract}

\maketitle

\section{INTRODUCTION}
\renewcommand{\theequation}{1.\arabic{equation}} \setcounter{equation}{0}

Einstein's theory of general relativity (GR) has stood as the most successful framework for understanding gravitational interactions for over a century, having consistently withstood a diverse array of experimental and observational tests. While most of these tests are in the regimes of weak gravitational fields and slow velocities of gravitational objects, the observations of gravitational waves (GWs) from compact binary coalescences enable us to probe gravity in extreme environments of strong gravitational fields and large velocities comparable to the speed of light. GW is one of the key predictions of GR and it was first directly detected by the LIGO, Virgo, and KAGRA scientific collaboration (LVK) \cite{LIGOScientific:2016aoc, LIGOScientific:2016vlm, LIGOScientific:2016emj, LIGOScientific:2016vbw}. By the end of the third observation run of LVK, about 90 GW events had been observed from the coalescence of compact binary systems \cite{LIGOScientific:2018mvr, LIGOScientific:2020ibl, KAGRA:2021vkt}. The O4 observation has been started and there will be more members added to the GW list without a doubt. These GW signals, including those from the merging of binary neutron stars, binary black holes, and binary neutron star and black holes, carry crucial information about the local spacetime properties of compact binaries in the strong field and highly dynamical spacetime region, helping us to test the predictions of GR.

There are two fundamental symmetries in GR: parity and Lorentz symmetries. It is well-documented early in the 1950s that the parity symmetry is broken in weak interaction \cite{Lee:1956qn, Wu:1957my}, whereas Lorentz symmetry has been confirmed with remarkable precision through particle experiments within the framework of the standard model of particle physics \cite{Mattingly:2005re, Kostelecky:2008ts}. It is natural to explore whether the parity and Lorentz symmetries could be broken in the gravitational sector. Theoretical attempts to integrate quantum physics with gravity, including string theory \cite{Kostelecky:1988zi, Kostelecky:1991ak}, loop quantum gravity \cite{Gambini:1998it}, and brane-world scenarios \cite{Burgess:2002tb}, etc., might lead to potential violations of these symmetries. There are also some modified theories of gravity with parity and Lorentz violations in the gravitational sectors, e.g., to mention a few, the Chern-Simons gravity \cite{Jackiw:2003pm, Alexander:2009tp}, chiral scalar-tensor theories \cite{Crisostomi:2017ugk}, parity-violating theory in the symmetric teleparallel gravity \cite{Conroy:2019ibo, Li:2022vtn, Li:2021mdp}, Ho\v{r}ava-Lifshitz gravity \cite{Horava:2009uw, Wang:2012fi, Zhu:2011xe, Zhu:2011yu}, Nieh-Yan modified teleparallel gravity \cite{Li:2021wij}, spatial covariant gravities \cite{Yu:2024drx, Hu:2024hzo, Gao:2019liu}. For an exhaustive list of modified gravities with parity and Lorentz violations, see Table.~I of Ref.~\cite{Zhu:2023rrx}. However, the constraints on the breakings of these symmetries within the gravitational sectors remain much less stringent, compared to those given through particle experiments \cite{Mattingly:2005re, Kostelecky:2008ts}.

When the parity and Lorentz symmetries in gravity are broken, they could induce possible derivations from the standard propagation properties of GWs in GR. Different mechanisms of parity and Lorentz violations may induce different effects in GW propagation. They could lead to frequency-dependent damping rate, nonlinear dispersion, and amplitude and velocity birefringences, which alter GW waveforms, see Refs.~\cite{Zhu:2023rrx, Zhao:2019xmm, Qiao:2019wsh} and reference therein \footnote{It is shown in Ref.~\cite{Hou:2024xbv} that with Lorentz violations, the extra scalar and vector types of GW polarizations can also be induced directly by the two tensorial GW modes. In this paper, we only focus on the parity and Lorentz-violating effects on the two tensorial GW modes.}. With specific derivations in GW propagation and waveforms from GR, one can obtain the constraints on the parity- and Lorentz-violating effects from GW data. This has enabled a lot of tests of parity and Lorentz symmetries by GW signals detected by LVK \cite{LIGOScientific:2019fpa, LIGOScientific:2020tif, LIGOScientific:2021sio, Wang:2020cub, Wu:2021ndf, Gong:2021jgg, Zhao:2022pun, Wang:2021gqm, Haegel:2022ymk, Wang:2025fhw, Gong:2023ffb, Zhu:2022uoq, Niu:2022yhr, Zhao:2019szi}. 

On the other hand, with the successful operation of LVK, it is confident to build more powerful GW detectors. The third-generation ground-based GW detectors, were composed for a few years, e.g., the Einstein Telescope (ET) in Europe \cite{Branchesi:2023mws} and the Cosmic Explorer (CE) in the United States \cite{Evans:2021gyd}. Compared to LVK, the ability of ET and CE is much better, and are expected to improve the detect range even to redshift $z=10$ for binary neutron stars and to $z=100$ for binary black holes with $30 M_{\odot}$ \cite{Evans:2021gyd}. Concurrently, several space-based GW detectors, such as LISA \cite{Robson:2018ifk, LISACosmologyWorkingGroup:2022jok}, Taiji \cite{Ruan:2018tsw, Wu:2018clg, Hu:2017mde}, and TianQin \cite{Liu:2020eko, Wang:2019ryf, TianQin:2015yph, Luo:2020bls} (see also Ref.~\cite{Gong:2021gvw}) are designed to explore the low-frequency GWs. These forthcoming GW detectors are expected to play a pivotal role in further propelling the frontier of gravitational physics.

In this paper, we investigate the prospects of constraining parity and Lorentz violations in gravity by using the simulated GW signals with the next generation of
ground-based GW detectors (ET and CE) and future space-based GW detectors (LISA, Taiji, and TianQin). For our purpose, we adopt a systematic parametric framework for characterizing possible derivations of GW propagation and modified waveform with parity- and Lorentz-violating effects \cite{Zhu:2023rrx, Zhao:2019xmm}. It is shown in Refs.~\cite{Zhu:2023rrx, Zhao:2019xmm} that this parametrization provides a general framework for studying the GW propagation of possible modifications caused by various modified gravitational theories, which has been used to study the constraints on parity and Lorentz violation from the data of compact binary merging events detected by LVK \cite{Zhu:2023rrx} and the effects in the primordial GWs \cite{Li:2024fxy}. For other alternative parametrized frameworks, see Refs. \cite{Nishizawa:2017nef, Ezquiaga:2021ler, Tahura:2018zuq, Saltas:2014dha, Nishizawa:2017nef}.

For ground-based GW detectors, we consider a network consisting of two GW detectors, ET and CE, and for space-based GW detectors, we use two different networks, LISA+Taiji and LISA+TianQin, respectively. We inject several typical GW signals from merging binary neutron stars and binary black holes into the specific GW detector networks. Then we perform the Bayesian analysis on the modified waveforms with parity- and Lorentz-violating effects to obtain the prospects for constraining parity and Lorentz violations in gravity.

This paper is organized as follows. In Sec. \ref{sec:2}, we introduce the parametrization for characterizing the parity- and Loentz-violating effects in GW propagation and the corresponding modified waveforms. In Sec. \ref{sec:3}, the brief introduction of Bayesian analysis and the settings of GW detectors used in our work are presented. In Sec. \ref{sec:4}, the main results of constraining the parity- and Lorentz-violating effects are presented and discussed. We summarize the results and conclusions in Sec. \ref{sec:5}. Throughout this paper, we adopt the units $c=G=1$ and the metric convention is chosen as $(-, +, +, +)$, and Greek indices $(\mu, \nu, \cdots)$ run over 0, 1, 2, 3 and Latin indices $(i, j, k, · · · )$ run over 1, 2, 3.

\section{MODIFIED WAVEFORM OF GWS WITH PARITY- AND LORENTZ-VIOLATING EFFECTS}\label{sec:2}
\renewcommand{\theequation}{2.\arabic{equation}} \setcounter{equation}{0}

In this section, we present a brief introduction of the modified waveforms of GWs with parity and Lorentz-violating effects, by adopting the systematic parametric framework described in Refs.~\cite{Zhu:2023rrx, Zhao:2019xmm}. Most of the expressions and results presented in this section can also be found in Refs.~\cite{Zhu:2023rrx, Zhao:2019xmm}.

We first introduce the propagation of GWs with parity- and Lorentz-violating effect. In a homogeneous and isotropic universe, GW is considered the perturbation of metric which is written as 
\begin{equation}
d s^2=a^2(\tau)\left[-d \tau^2+\left(\delta_{i j}+h_{i j}\right) d x^i d x^j\right],
\end{equation}
where $a(\tau)$ is the scale factor of the expanding universe and the present expansion factor is set to $a_{0}=1$. $\tau$ represents the conformal time and it can be transformed to cosmic time $t$ by the equation $dt = a(\tau)d\tau$. 
We focus on the modes $h_{ij}$ which are transverse and traceless,
\begin{equation}
    \partial^{i}h_{ij}=0=h^{i}_{i}.
\end{equation}
One can expand $h_{ij}$ over the spatial Fourier harmonics as
\begin{equation}
    h_{ij}(\tau,x^{i})=\sum_{A=R,L}\int\frac{d^{3}x}{(2\pi)^{3}}h_A(\tau,k^{i})e^{ik_{i}x^{i}}e^{A}_{ij}(k^{i}),
\end{equation}
where $e^{A}_{ij}$ represent the circular polarization tensors which can be transform by $\epsilon^{ijk}n_{i}e^{A}_{kl}=i\rho_{A}e^{jA_{l}}$ with $\rho_{R}=1$ and $\rho_{L}=-1$.

To parametrize the parity- and Lorentz-violating effects on the GW propagation, one can describe the modified propagation equations of the two GW modes by four extra parameters \cite{Zhu:2023rrx, Zhao:2019xmm},
\begin{equation}
    h_A^{\prime \prime}+\left(2+\bar{\nu}+\nu_A\right) \mathcal{H} h_A^{\prime}+\left(1+\bar{\mu}+\mu_A\right) k^2 h_A=0.\lb{2.4}
\end{equation}
Here, a prime represents the derivative for the conformal time $\tau$ and $\mathcal{H}=a'/a$. Here, we would like to address some remarks about the validity and the underlined assumptions of Eq. (\ref{2.4}). This equation is written under the linear approximation, which implies we have ignored all the high-order metric perturbations on the propagation of GWs. We also restrict us to the transverse and traceless modes of GWs for two reasons. Theoretically, the metric perturbations $h_{ij}$ only contains two degenerate traceless and transverse tensor modes in GR, but when the parity- and Lorentz-violating modifications are included, depending on specific types of the parity and Lorentz violations, $h_{ij}$ may contain extra modes, for example, the scalar or vector modes. These extra modes decouple from the two traceless and transverse tensor modes under the linear approximation and thus do not affect the propagation of the two tensorial modes. Observationally, all the GW signals detected by LIGO/Virgo/KAGRA detectors can be well described by the two tensorial polarizations, with no conclusive evidence for scalar or vector modes \cite{KAGRA:2021vkt}. In addition, the Lorentz- and parity-violating terms included in this equation are assumed to be subdominant compared to the leading-order GR contributions \cite{Zhao:2019xmm}, which constrains parameter ranges to avoid ghosts or instabilities.

The four parameters, $\bar{\nu}$, $\bar{\mu}$, $\nu_A$, and $\mu_A$, are introduced to describe deviations from the standard GW propagation in GR, which can arise from specific modified theories of gravity. These parameters capture three distinct classes of effects: (1) Frequency-independent modifications associated with $\bar{\nu}$ and $\bar{\mu}$, which account for changes in the propagation speed of GWs and additional friction effects. (2) Parity-violating effects described by $\nu_A$ and $\mu_A$, which induce birefringence phenomena, including amplitude and velocity birefringences, in GW propagation. (3) Lorentz-violating effects, also governed by $\bar{\nu}$ and $\bar{\mu}$, which introduce frequency-dependent damping and nonlinear dispersion of GWs. Forecasts for constraints on the frequency-independent and frequency-dependent components of GW friction achievable by upcoming GW detectors are discussed in Refs. \cite{Lin:2024pkr, Zhang:2024rel}, while constrains on birefringences \cite{Wang:2020cub, Califano:2023aji} and Lorentz-violating dispersion relation have been carried out in Ref.~\cite{Mirshekari:2011yq}. In this work, we focus on the latter two frequency-dependent categories.

\subsection{Modified GW waveforms with parity-violating effects}

The parity-violating effects are represented by the parameters $\nu_A$ and $\mu_A$. The parameter $\nu_A$ influences the damping rates of the left- and right-hand circular polarizations of GWs, a phenomenon known as amplitude birefringence. This means that during propagation, the amplitude of the left-hand mode decreases (or increases), while the amplitude of the right-hand mode increases (or decreases). On the other hand, the parameter $\mu_A$ affects the velocities of the left- and right-hand circular polarizations, leading to velocity birefringence. Consequently, the left- and right-handed modes will not arrive simultaneously. We can express the frequency-dependent parameters $\nu_A$ and $\mu_A$ as \cite{Zhu:2023rrx, Zhao:2019xmm}
\begin{eqnarray}
\mathcal{H} \nu_{\mathrm{A}} &=&\left[\rho_{\mathrm{A}} \alpha_{\nu}(\tau)\left(k / a M_{\mathrm{PV}}\right)^{\beta_{\nu}}\right]^{\prime}, \\
\mu_{\mathrm{A}}&=&\rho_{\mathrm{A}} \alpha_{\mu}(\tau)\left(k / a M_{\mathrm{PV}}\right)^{\beta_{\mu}},
\end{eqnarray}
where $\alpha_\nu$, $\alpha_\mu$ are arbitrary functions of time, $\beta_\nu$, $\beta_\mu$ are arbitrary odd numbers, and $M_{\rm PV}$ are the energy scale of the parity violations. We analyze the simulated GW events that are in the local Universe, thus these two equations can be considered as constant. The previous work \cite{Zhu:2023rrx} summarizes the various values of $\mathcal{H} \nu_{\rm A}, \mu_{\rm A}, \beta_{\nu}$, and $\beta_{\mu}$ corresponding to different modified theories of gravity. 

The modified GW waveform with the parity-violating effects can be written based on the GR-based form with the stationary phase approximation (SPA)  \cite{Zhu:2023rrx, Zhao:2019xmm}
\begin{equation}
\tilde{h}_A(f)=\tilde{h}_A^{\mathrm{GR}} e^{\rho_A \delta h_1} e^{i\left(\rho_A \delta \Psi_1\right)},
\label{pv_ha}
\end{equation}
where $\tilde{h}_A^{\mathrm{GR}}$ is the GW waveform of GR, whose detail of the waveform is discussed in Ref.~\cite{Zhao:2019xmm}. The amplitude correction, induced by the parameter $\nu_A$, can be expressed as $\delta h_1 = A_{\nu}(\pi f)^{\beta_{\nu}}$. The parameter $\mu_A$ induces the phase correction $\delta \Psi_1 = A_{\mu}(\pi f)^{\beta_{\nu}+1}$ for $\beta_{\mu} \neq -1$, and $\delta \Psi_{1} = A_{\mu} \ln u$ for $\beta_{\mu}=-1$. The explicit form of the parameters $A_{\nu}$ and $A_{\mu}$ can be written as
\bqn
 A_\nu&=&\frac{1}{2}\left(\frac{2}{M_{\mathrm{PV}}}\right)^{\beta_\nu}\left[\alpha_\nu\left(\tau_0\right)-\alpha_\nu\left(\tau_e\right)(1+z)^{\beta_\nu}\right], \nb\\
A_\mu&=&\frac{\left(2 / M_{\mathrm{PV}}\right)^{\beta_\mu}}{\Theta\left(\beta_\mu+1\right)} \int_0^z \frac{\alpha_\mu\left(1+z^{\prime}\right)^{\beta_\mu}}{H_0 \sqrt{\Omega_m\left(1+z^{\prime}\right)^3+\Omega_{\Lambda}}} d z^{\prime}, \nb \\
\label{eq:pv_anu_amu}
\eqn
where $\tau_{e} (\tau_{0})$ is the emitted (arrival) time of GWs, $z=1/a(\tau_{e})-1$ is the redshift of GW event, $u=\pi \mathcal{M} f$, $\mathcal{M}$ is the measured chirp mass of the compact binary system, $f$ is the frequency of GW detected by the GW detectors, and the function $\Theta(1+x)=1+x$ for $x \neq -1$ and $\Theta(1+x)=1$ for $x = -1$. In this work, the cosmology parameter we adopt are $\Omega_m=0.315$, $\Omega_{\Lambda}=0.685$, and $H_0=67.4\; {\rm km}\;{\rm s}^{-1}\; {\rm Mpc}^{-1}$ \cite{Planck:2018vyg} \footnote{Here we use the Planck cosmological parameters for consistency with previous results presented in \cite{LIGOScientific:2019fpa, LIGOScientific:2021sio,Gong:2023ffb, Wang:2025fhw, Wang:2020cub,LIGOScientific:2020tif,Wu:2021ndf}. Using different sets of cosmological parameters can only slightly change the results presented in this paper.}.

\subsection{Modified GW waveforms with Lorentz-violating effects}

The parameters $\bar{\nu}$ and $\bar{\mu}$ express the Lorentz-violating effects. The frequency-dependent friction during the propagation of GWs is induced by the parameter $\bar{\nu}$ and the non-linear dispersion relation of GWs is induced by $\bar{\mu}$. Given that both $\bar{\nu}$ and $\bar{\mu}$ are frequency-dependent, they can be parameterized as follows \cite{Zhu:2023rrx, Zhao:2019xmm}
\begin{eqnarray}
\mathcal{H} \bar{\nu} &=&\left[\alpha_{\bar{\nu}}(\tau)\left(k / a M_{\mathrm{LV}}\right)^{\beta_{\bar{\nu}}}\right]', \\
\bar{\mu}&=&\alpha_{\bar{\mu}}(\tau)\left(k / a M_{\mathrm{LV}}\right)^{\beta_{\bar{\mu}}}, 
\end{eqnarray}
where $\alpha_{\bar \nu}$, $\alpha_{\bar \mu}$ are arbitrary functions of time, $\beta_{\bar \nu}$, $\beta_{\bar \mu}$ are arbitrary even numbers. In the local universe, we consider them as constants. And $M_{\rm LV}$ represents the energy scale of Lorentz violation. The previous work \cite{Zhu:2023rrx} summarizes the various values of $\mathcal{H} \bar{\nu}, \bar{\mu}, \beta_{\bar{\nu}}$, and $\beta_{\bar{\mu}}$ corresponding to different modified theories of the gravity.

These two frequency-dependent parameters affect different aspects of GW during its propagation. For the parameter $\bar{\nu}$, it means the different frequencies of GWs will induce different damping rates, which are expressed as amplitude modulation in the waveform of GW. For the parameter $\bar{\mu}$, it affects phase velocity and induces a modified phase term in the waveform of GW. The modified waveform of GW with Lorentz-violating effects can be written by \cite{Zhu:2023rrx, Zhao:2019xmm} (see also Ref. \cite{Mirshekari:2011yq} for the modified waveforms for Lorentz-violating dispersion relations)
\begin{equation}
\tilde{h}_A(f)=\tilde{h}_A^{\mathrm{GR}}(f) e^{\delta h_2} e^{i \delta \Psi_2},
\label{lv_ha}
\end{equation}
where the modified amplitude term $\delta h_2 = -A_{\bar{\nu}} (\pi f)^{\beta_{\bar{\nu}}}$, the modified phase term $\delta \Psi_2 = A_{\bar{\mu}} (\pi f)^{\beta_{\bar{\mu}}+1}$ for $\beta_{\bar{\mu}} \neq -1$ and $\delta \Psi_2 = A_{\bar{\mu}} \ln u$ for $\beta_{\bar{\mu}} = -1$. The specific expressions for $A_{\bar{\nu}}$ and $A_{\bar{\mu}}$ are give by
\begin{eqnarray}
A_{\bar{\nu}} &=& \frac{1}{2} \left(\frac{2}{M_{\rm LV}}\right)^{\beta_{\bar \nu}}\Big[\alpha_{\bar \nu}(\tau_0) - \alpha_{\bar \nu}(\tau_e) (1+z)^{\beta_{\bar \nu}}\Big], \nb\\
A_{\bar{\mu}} &=& \frac{(2/M_{\rm LV})^{\beta_{\bar \mu}}}{\Theta(\beta_{\bar \mu}+1)} \int_{0}^{z} \frac{\alpha_{\bar \mu} (1+z')^{\beta_{\bar \mu}}}{H_0 \sqrt{\Omega_m(1+z')^3 +\Omega_\Lambda}}dz'. \nb\\ \label{eq:lv_anu_amu}
\end{eqnarray}

\section{Bayesian analysis on injected GW signals into the future ground-based GW detectors and space-based GW detectors}\label{sec:3}
\renewcommand{\theequation}{3.\arabic{equation}} \setcounter{equation}{0}

\begin{table*}
\caption{\label{tab:detector list}%
Configuration of six GW detectors. We consider two ground-based GW detectors, ET and CE, along with three space-based GW detectors, LISA, Taiji, and TianQin.}
\begin{ruledtabular}
\begin{tabular}{cccccc}
Detector & Configuration & $f_{\rm lower}$ [Hz] & $f_{\rm upper}$ [Hz]  & Arm length [km] & Reference\\
\hline
CE & Rightangle & 5 & 5000 & 20/40 & Ref. \cite{Reitze:2019iox}\\
ET & Triangle & 1 & 10000 & 10 & Ref. \cite{Hild:2010id}\\
TianQin & Triangle &  0.0001 & 1 & $17 \times 10^{4}$ & Ref. \cite{TianQin:2020hid}\\
LISA & Triangle &  0.0001 &  0.1 & $2.5 \times 10^{6}$ & Ref. \cite{Liu:2023qap}\\
Taiji & Triangle &  0.0001 &  0.1 & $3 \times 10^{6}$ & Ref. \cite{Liu:2023qap}\\
\end{tabular}
\end{ruledtabular}
\end{table*}

\begin{table}
\caption{\label{tab:event list}%
The main parameters of the six simulated GW events.}
\begin{ruledtabular}
\begin{tabular}{cccc}
Event & $m_1 (M_\odot)$ &  $m_2 (M_\odot)$ & Redshift $z$ \\
\hline
GW170817-like & 1.46 & 1.27 & 0.01\\
GW150914-like & 35.6 & 30.6 & 0.09 \\
GW190521-like & 98.4 & 57.2 & 0.56 \\
event1 & $10^4$ & $10^4$ & 1 \\
evnet2 & $10^5$ & $10^5$ & 5 \\
event3 & $10^6$ & $10^6$ & 10\\
\end{tabular}
\end{ruledtabular}
\end{table}

\subsection{Bayesian inference for GW data}

In this subsection, we introduce the Bayesian inference to analyze the simulated GW signals for constraining the parity and Lorentz violation with the modified waveform of GW given in Eq.~(\ref{pv_ha}) and Eq.~(\ref{lv_ha}).

Bayesian inference is rapidly emerging as a valuable tool in cosmology and GW astronomy. In research on GWs, it is mainly used to infer the astrophysical properties of GW sources. In this paper, we focus on inferring the modified waveform parameters of GWs using simulated GW signals with future GW detectors. The future ground- and space-based GW detectors are considered.

In Bayes's theorem, the model-dependent posterior distribution for parameters $\theta$ can be expressed as
\bqn
P(\vec{\theta}|d,H) = \frac{P(d|\vec{\theta},H) P(\vec{\theta}|H)}{P(d|H)},
\eqn
where $d_i$ is the simulated GW data and the parameters $\vec{\theta}$ represent the model parameters that describe the modified GW waveforms with parity- or Lorentz-violating effects. In this equation, the posterior probability distribution $P(\vec{\theta}|d, H)$ is based on the data $d$ and the waveform model $H$. $P(d|\vec{\theta}, H)$ is the likelihood, which stands for the probability of observing the data given the parameters $\vec{\theta}$ and model $H$. $P(\vec{\theta}|H)$ is the prior distribution, which means the probability of the parameters $\vec{\theta}$. And $P(d|H)$ is the evidence representing the probability of observing the data with model $H$, which can be written by 
\bqn
P(d|H) \equiv \int d\vec{\theta} P(d|\vec{\theta},H) P(\vec{\theta}|H).
\eqn
This method offers a statistical analysis to constrain the parameters of the modified waveform of GW which express the parity- and Lorentz-violating effects with GW signals.

To analyze GW signals, the matched filtering method is in general used to extract the GW signals from the observed data with noises because the real GW signal is extremely weak. The likelihood function for the matched filtering method is written as 
\begin{equation} \label{eq:likelihood}
P(d|{\vec{\theta}}, H) \propto \prod_{i=1}^{n} \exp\left(-\frac{1}{2}\langle d_i-h(\vec{\theta})|{d_i}-{h}(\vec{\theta})\rangle\right),
\end{equation}
where $i$ represents the different GW detectors and the GW strain $h(\vec{\theta})$ is predicted by the waveform model $H$. $\langle h_1|h_2 \rangle$ is the noise-weighted inner product expressed as

\begin{equation} \label{eq:inner_product}
\langle h_1|h_2 \rangle = 4\, \text{Re} \left[\int_0^\infty \frac{h_1(f) h_2^*(f)}{S_n(f)} \,df\right],
\end{equation}
where star superscript stands for complex conjugation and $S_n(f)$ denotes the power spectral density (PSD) of the GW detectors' noise. The PSD data in our analysis contain the design sensitivities of ET, CE, LISA, Taiji, and TianQin \cite{Liu:2023qap, TianQin:2020hid, LISA:2017pwj, Reitze:2019iox, Evans:2021gyd, Hild:2010id}.

The modified GW waveforms of the circular polarization modes can be written as \cite{Zhu:2023rrx, Zhao:2019xmm}
\begin{eqnarray}
 \tilde h_A(f) = \tilde h_A^{\rm GR}(f) e^{ \rho_A \delta h_1 +\delta h_2} e^{i  (\rho_A \delta \Psi_1 + \delta \Psi_2)},\label{waveforms}
 \end{eqnarray}
which contains both parity- and Lorentz-violating effects. One can transform circular polarization modes $\tilde h_{\rm R}$ and $\tilde h_{\rm L}$ to the polarization modes $\tilde h_{+}$ and $\tilde h_{\times}$ via $\tilde h_{+} = \frac{\tilde h_{\rm L} + \tilde h_{\rm R}}{\sqrt{2}}$ and $\tilde h_{\times} = \frac{\tilde h_{\rm L} - \tilde h_{\rm R}}{\sqrt{2}i}$. 
 
In the above-modified waveforms, the parity- and Lorentz-violating effects are labeled by the parameters $A_\nu$, $A_\mu$, $A_{\bar \nu}$, and $A_{\bar \mu}$. These four parameters are also the extra parameters along with the GR parameters we analyzed in the Bayesian inference. One then can perform the Bayesian inference with the above-modified waveforms on the simulated GW signals, which are supposed to be detected by different sets of GW detectors. By considering a series of GW signals through the Bayesian inference, one can infer posterior distributions of the parameters $A_\nu$, $A_\mu$, $A_{\bar \nu}$, and $A_{\bar \mu}$ for each signal.

In the following subsections, we describe the details of the analysis with the ground-based GW detectors and space-based GW detectors respectively.

\subsection{Analysis with ground-based GW detectors}

For the third generation of ground-based GW detectors, we focus on the capabilities of a ground-based GW detector network consisting of ET and CE, for constraining the parity and Lorentz violations. For ET, it will be built in a triangle shape in Europe with a 10\;km arm length. The Cosmic Explorer design includes two facilities: one with a 40\;km arm length in the United States and the other with a 20\;km arm length in the Australia. Both facilities will feature an L-shaped GW detector. The location of ET and CE (the United States) we used in simulating GW signals are ($43.70 \degree$, $10.42 \degree$) and (-$33.29 \degree$, $149.09 \degree$) separately. We summarize the other configurations of GW detectors in Table \ref{tab:detector list}. For sensitivity profiles, CE employs the CE-strain sensitivity curve \cite{Srivastava:2022slt, Evans:2021gyd}, and ET utilizes the ET-D curve \cite{Hild:2010id} for assessing strain sensitivity both in amplitude and spectral density.

We then inject several typical GW signals into ET and CE by using the open source package \texttt{BILBY} \cite{Ashton:2018jfp,Romero-Shaw:2020owr}. We select three representative GW events (GW150914-like, GW170817-like, and GW190521-like) by using the parameters of these GW event in GWTC-3 \cite{KAGRA:2021vkt} as the parameters of the injected signals. One of the main reasons for these choices is for the later convenience of comparison with the results from these specific GW events \cite{KAGRA:2021vkt}. For the binary black hole events (GW150914-like and GW190521-like), we adopt the waveform template \texttt{IMRPhenomPv2} \cite{Hannam:2013oca, Khan:2015jqa}, while for binary neutron star (GW170817-like), we use \texttt{IMRPhenomPv2\_NRTidal} \cite{Dietrich:2017aum, Dietrich:2018uni, Dietrich:2019kaq}, as the waveforms in the generation of the simulated GW signals. The properties of three simulated GW events GW150914-like, GW170817-like, and GW190521-like are shown in Table. \ref{tab:event list}.

We consider the modified waveforms of GW in Eq.~(\ref{waveforms}) with different values of ($\beta_{\nu}$, $\beta_{\mu}$) and ($\beta_{\bar \nu}$, $\beta_{\bar \mu}$) separately, which represent parity- and Lorentz-violating effects respectively. For parity-violating effects, we consider four different values of $\beta_{\nu}$ and $\beta_\mu$, in which  $\beta_{\nu}=1$ denotes amplitude birefringence, and $\beta_{\mu}=-1, 1, 3$ represent velocity birefringence. For Lorentz-violating effects, we consider four different values of $\beta_{\bar \nu}$ and $\beta_{\bar \mu}$, in which  $\beta_{\bar \nu}=2$ corresponds to  frequency-dependent damping effect \cite{Zhang:2024rel}, and $\beta_{\bar \mu}=-2, 2, 4$ represent modified  dispersion relations of GWs. Compared to the six cases considered in a previous analysis with LVK data \cite{Zhu:2023rrx}, here we additionally consider two more cases $\beta_{\mu}=1$ and $\beta_{\bar \mu}=-2$. In the Bayesioan analysis, the modified GW waveforms, containing parity violation as described in Eq.~(\ref{eq:pv_anu_amu}) and Lorentz violation as detailed in Eq.~(\ref{lv_ha}), are based on the GR waveform template available in LALSuite \cite{LALSuite2018}. We follow the simulated data using the template \texttt{IMRPhenomPv2} for GW150914-like and GW190521-like events, and employing the template \texttt{IMRPhenomPv2\_NRTidal} for GW170817-like event. Based on these templates of GR, we add the parity- and Lorentz-violating effects on the waveforms and test these 8 cases separately. We refer to Refs.~\cite{LIGOScientific:2018mvr, LIGOScientific:2020ibl, KAGRA:2021vkt, Zhu:2023rrx} to set the priors for the GR parameters in the modified waveforms. For parity- and Lorentz-violating parameters $A_\nu$, $A_\mu$, $A_{\bar \nu}$, and $A_{\bar \mu}$, their priors are based on $\delta \Psi_1 = A_{\mu}(\pi f)^{\beta_{\nu}+1}$ and $\delta h_2 = -A_{\bar{\nu}} (\pi f)^{\beta_{\bar{\nu}}}$, which are set to be uniformly distributed. Then in the part of parameter estimations, we employ the open source package \texttt{BILBY} \cite{Romero-Shaw:2020owr, Ashton:2018jfp} and the nested sampling method \texttt{dynesty} \cite{Speagle:2019ivv}.

\subsection{Analysis with space-based GW detectors}

We now turn to space-based GW detectors, including LISA, Taiji, and TianQin. Specifically, we evaluate the capabilities of two GW detector networks in constraining parity and Lorentz violations: one composed of LISA and Taiji (LISA+Taiji) and another composed of LISA and TianQin (LISA+TianQin). These GW detectors are expected to begin operations in the 2030s. For our analysis, we adopt the sensitivity curves of LISA, Taiji, and TianQin as given in Refs.~\cite{Liu:2023qap, TianQin:2020hid}. The configurations of these GW detector networks are summarized in Table~\ref{tab:detector list}.

We then simulate the GW signals from three typical GW signals from the merging of supermassive binary black hole systems which are expected to be detected by LISA-Taiji and LISA-TianQin by using the open source package \texttt{BILBY} \cite{Ashton:2018jfp,Romero-Shaw:2020owr}. The information of these three simulated signals, denoted by event 1, event 2, and event 3, respectively, are summarized in Table \ref{tab:event list}. To generate the GW signals through \texttt{BILBY} \cite{Ashton:2018jfp,Romero-Shaw:2020owr}, we adopt the waveform template \texttt{IMRPhenomPv2}. 

The main parameters of the mass and redshift of GW signals are shown in Table \ref{tab:event list}, with other parameters of simulated GW signals being the same as GW signals of ground-based GW detectors. Similar to the analysis for the ground-based GW detectors, we then perform the Bayesian analysis with the modified waveforms of GW in Eq.~(\ref{waveforms}) for 8 different values of ($\beta_{\nu}$, $\beta_{\mu}$) and ($\beta_{\bar \nu}$, $\beta_{\bar \mu}$) separately. We present the main results from these analyses in the next section.

\section{Constraints on parity and Lorentz violations}\label{sec:4}
\renewcommand{\theequation}{4.\arabic{equation}} \setcounter{equation}{0}

\begin{figure*}
\centering
\includegraphics[width=4.4cm]{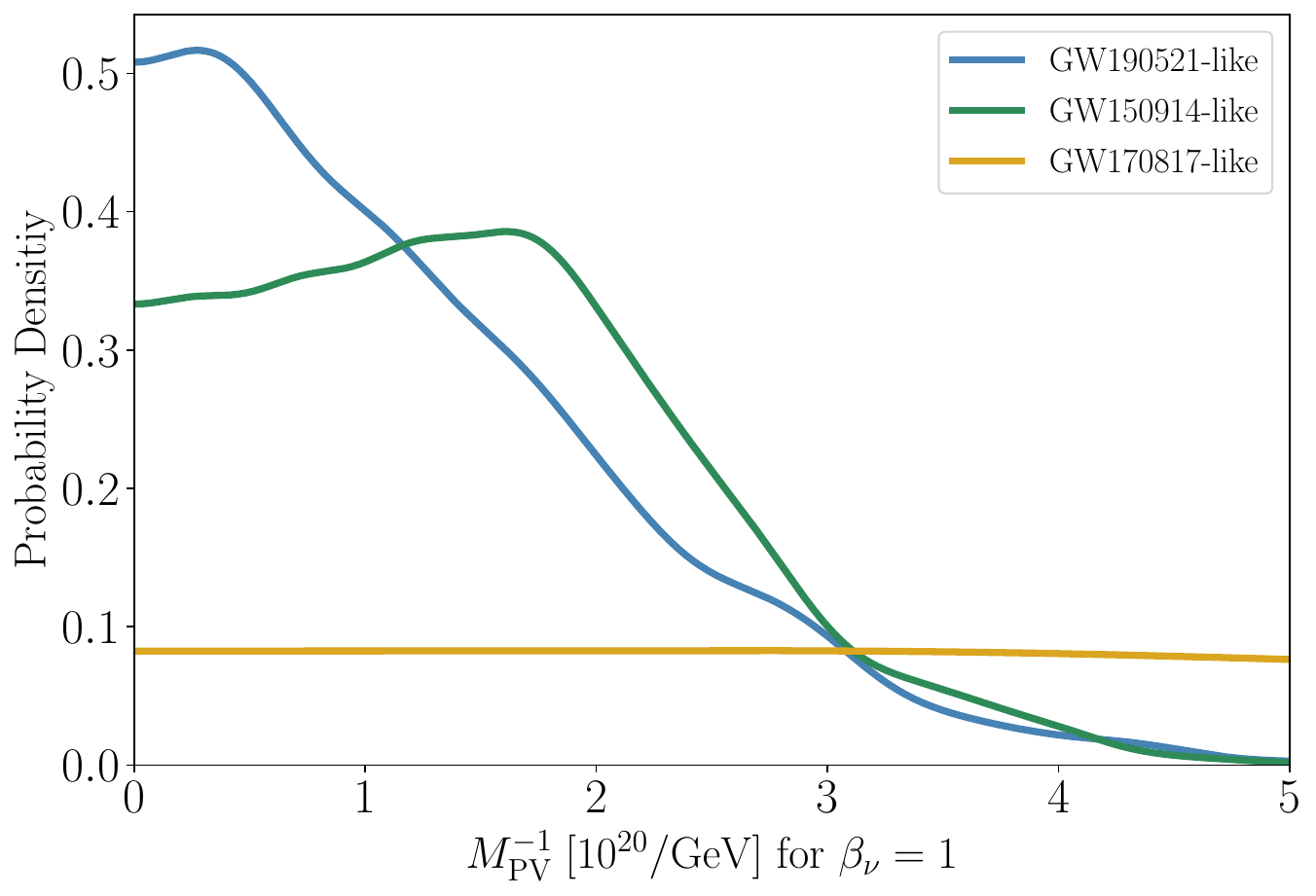}
\includegraphics[width=4.4cm]{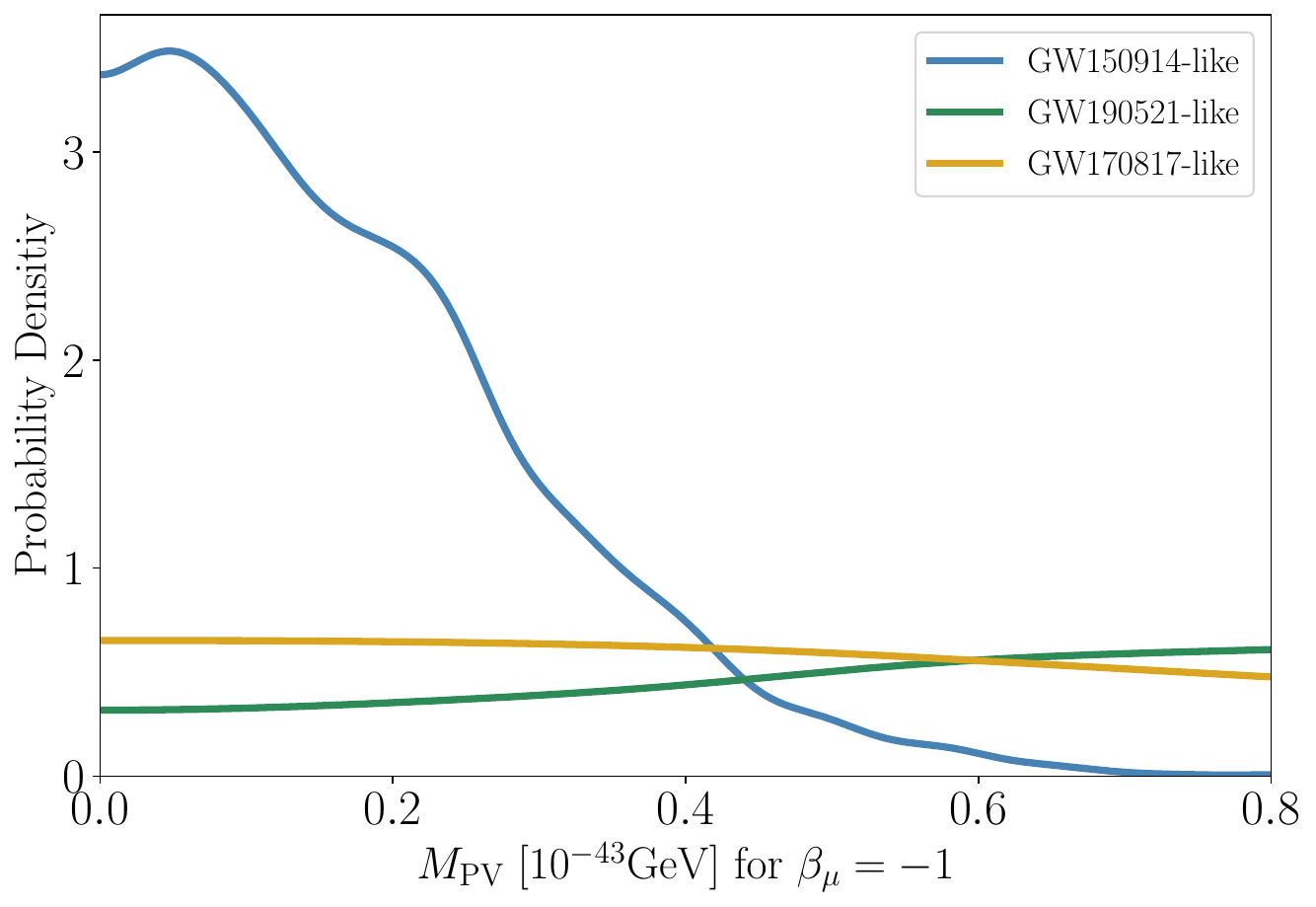}
\includegraphics[width=4.4cm]{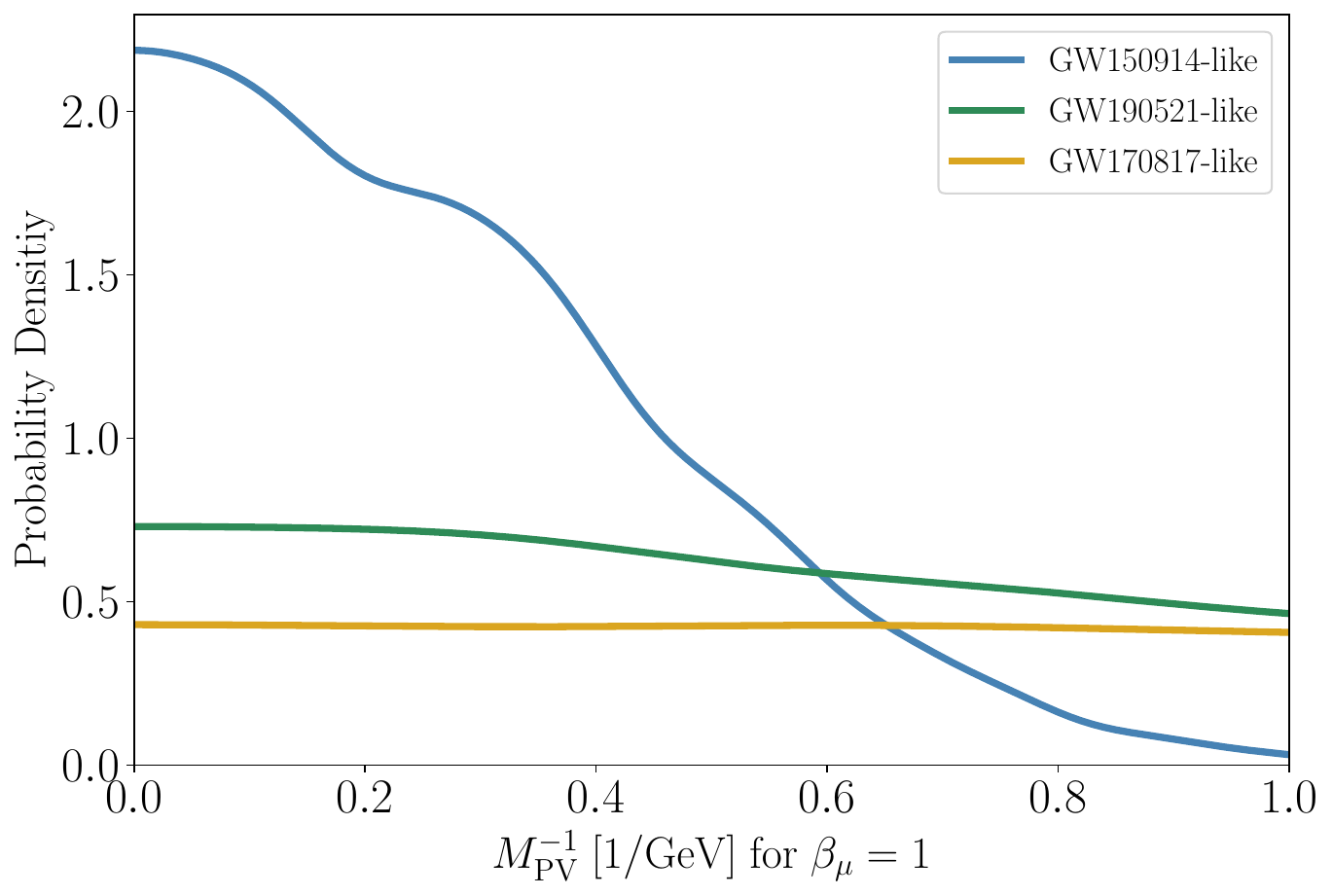}
\includegraphics[width=4.4cm]{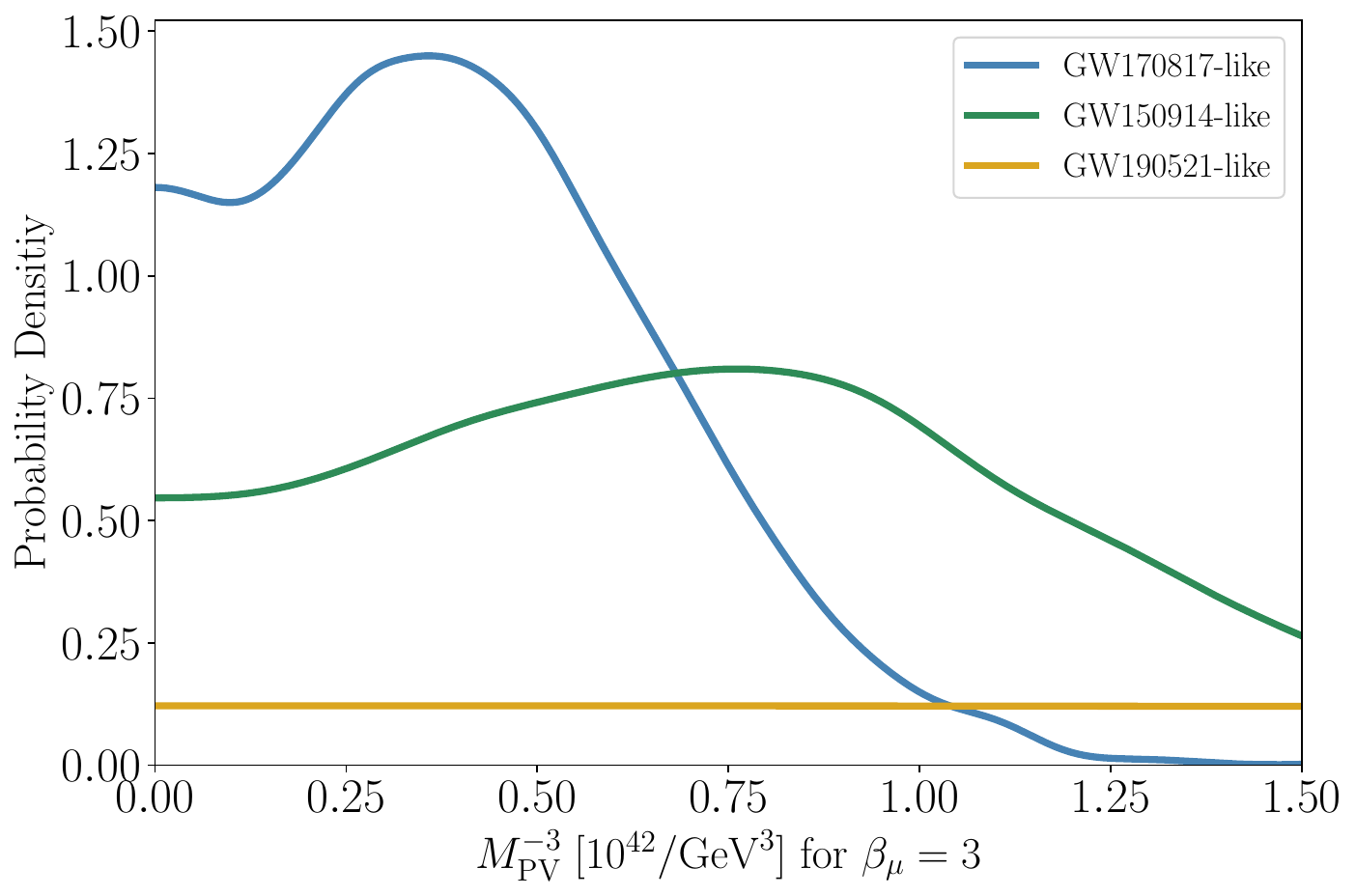}
\includegraphics[width=4.4cm]{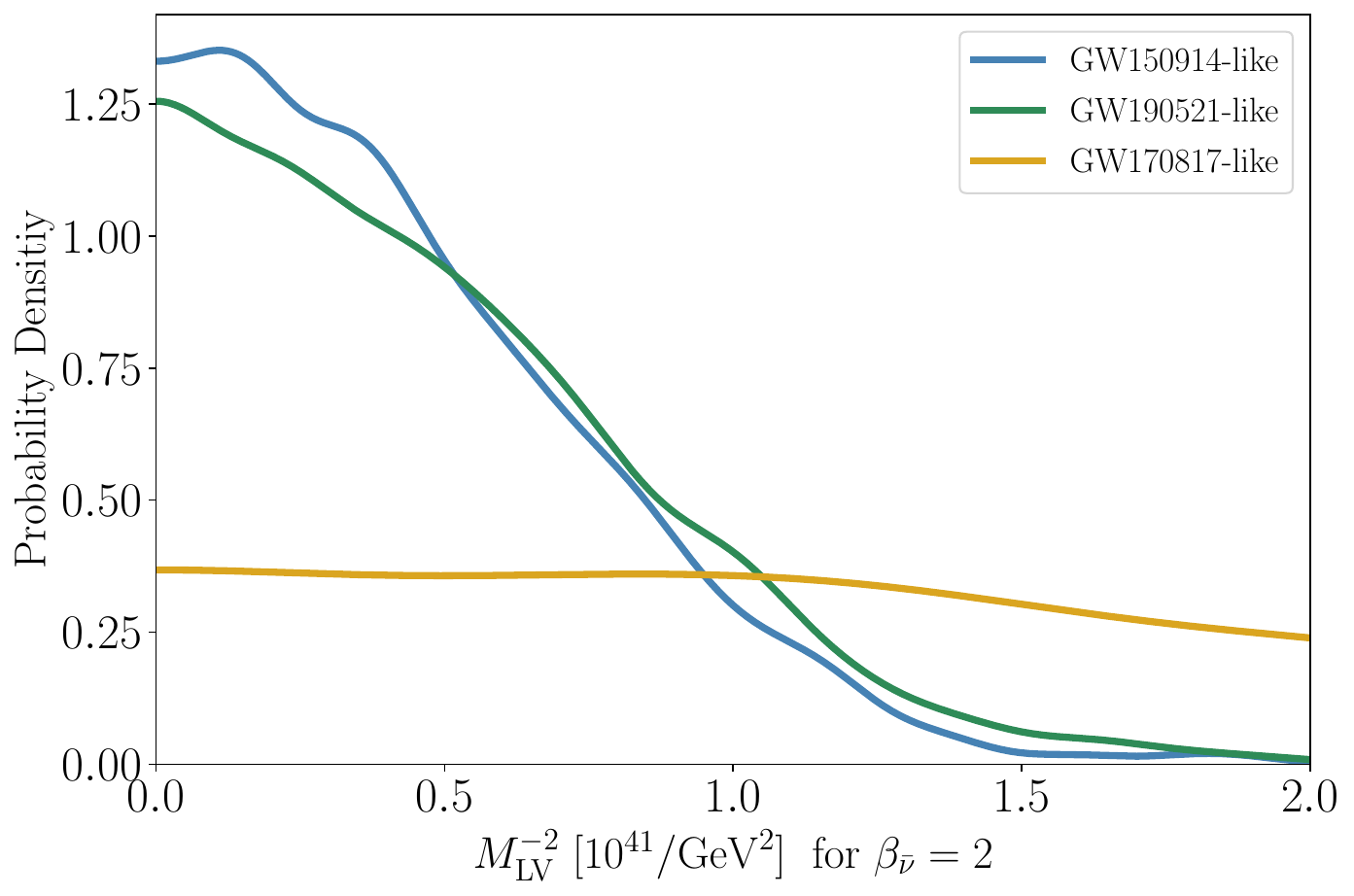}
\includegraphics[width=4.4cm]{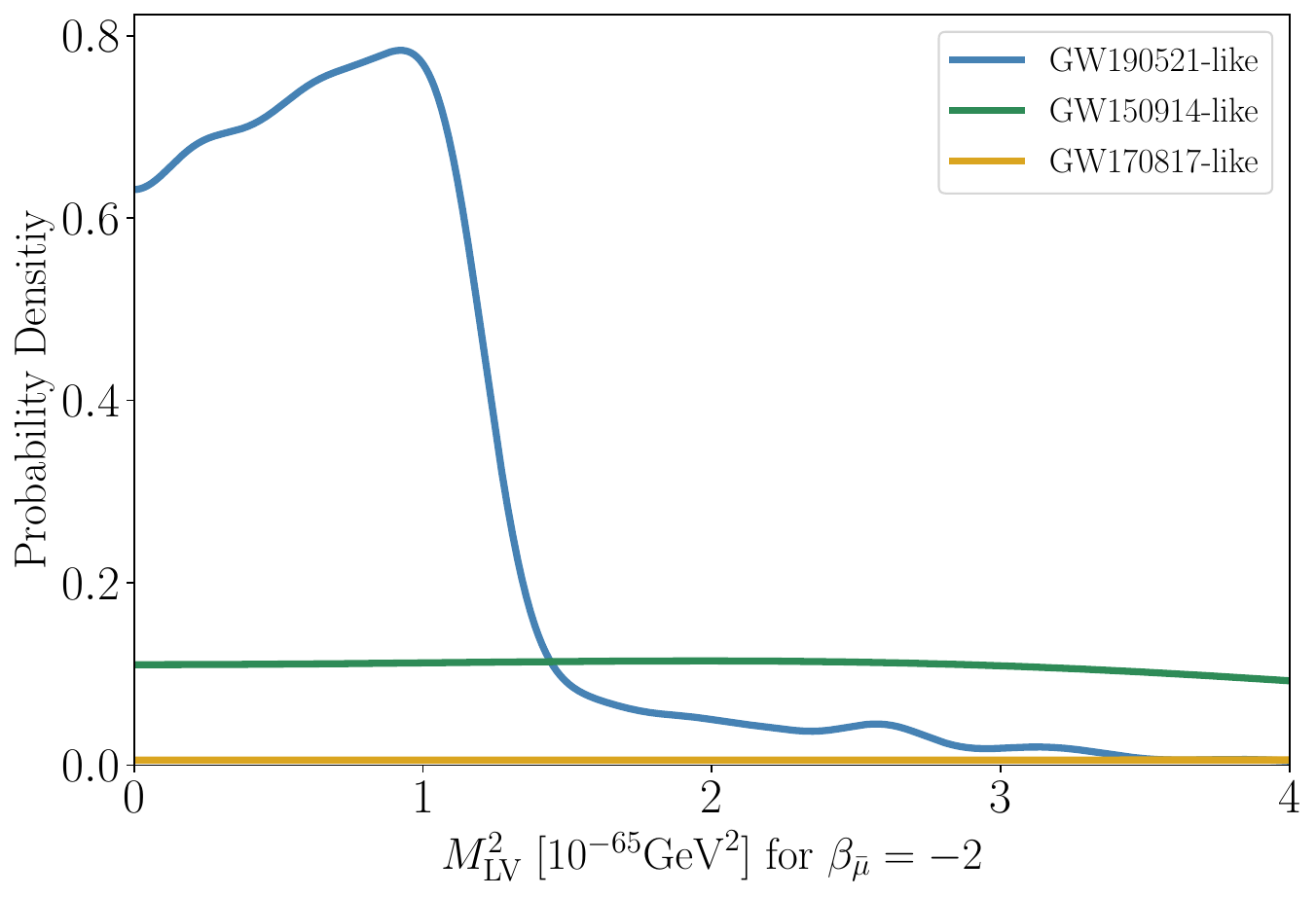}
\includegraphics[width=4.4cm]{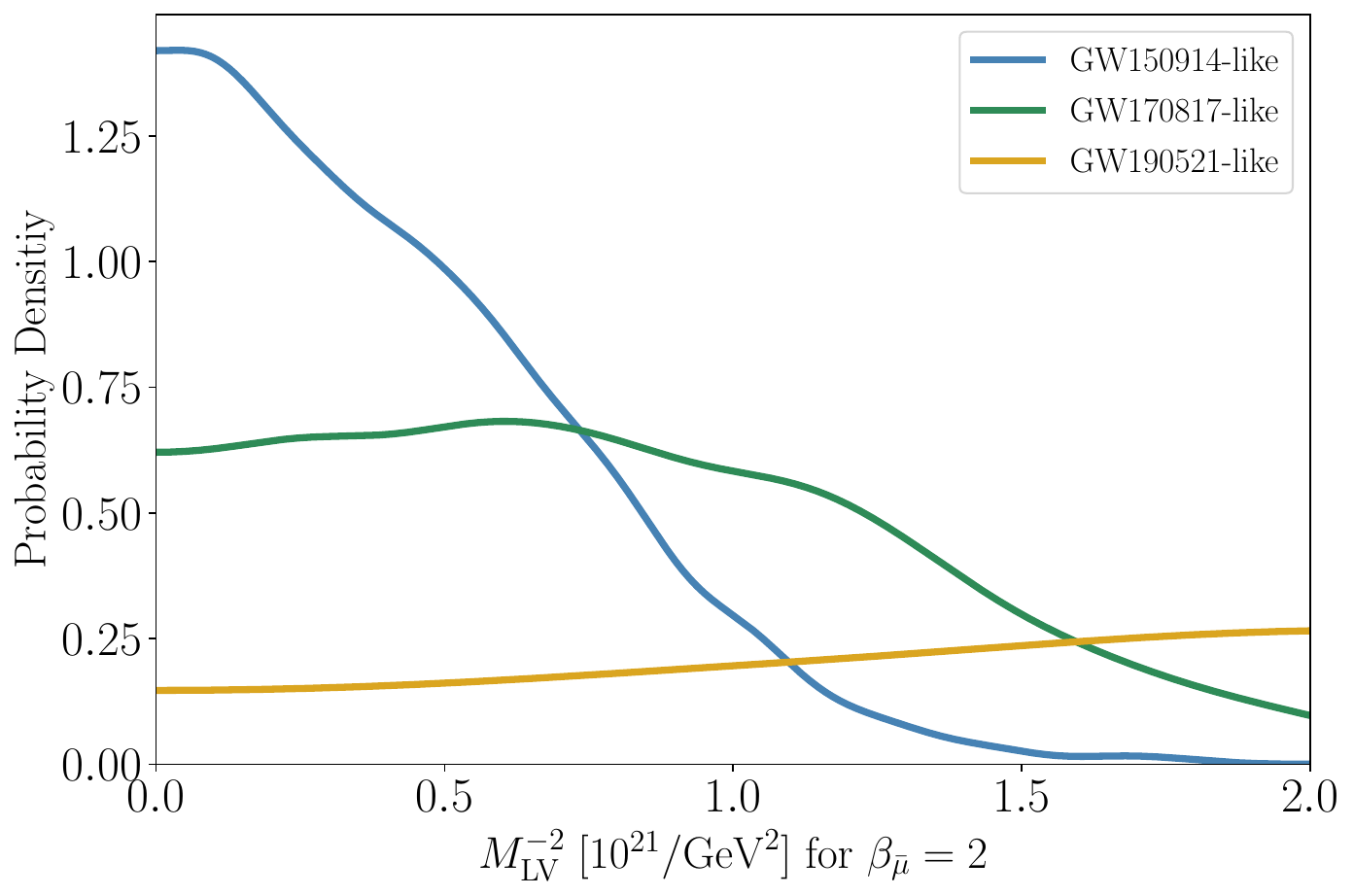}
\includegraphics[width=4.4cm]{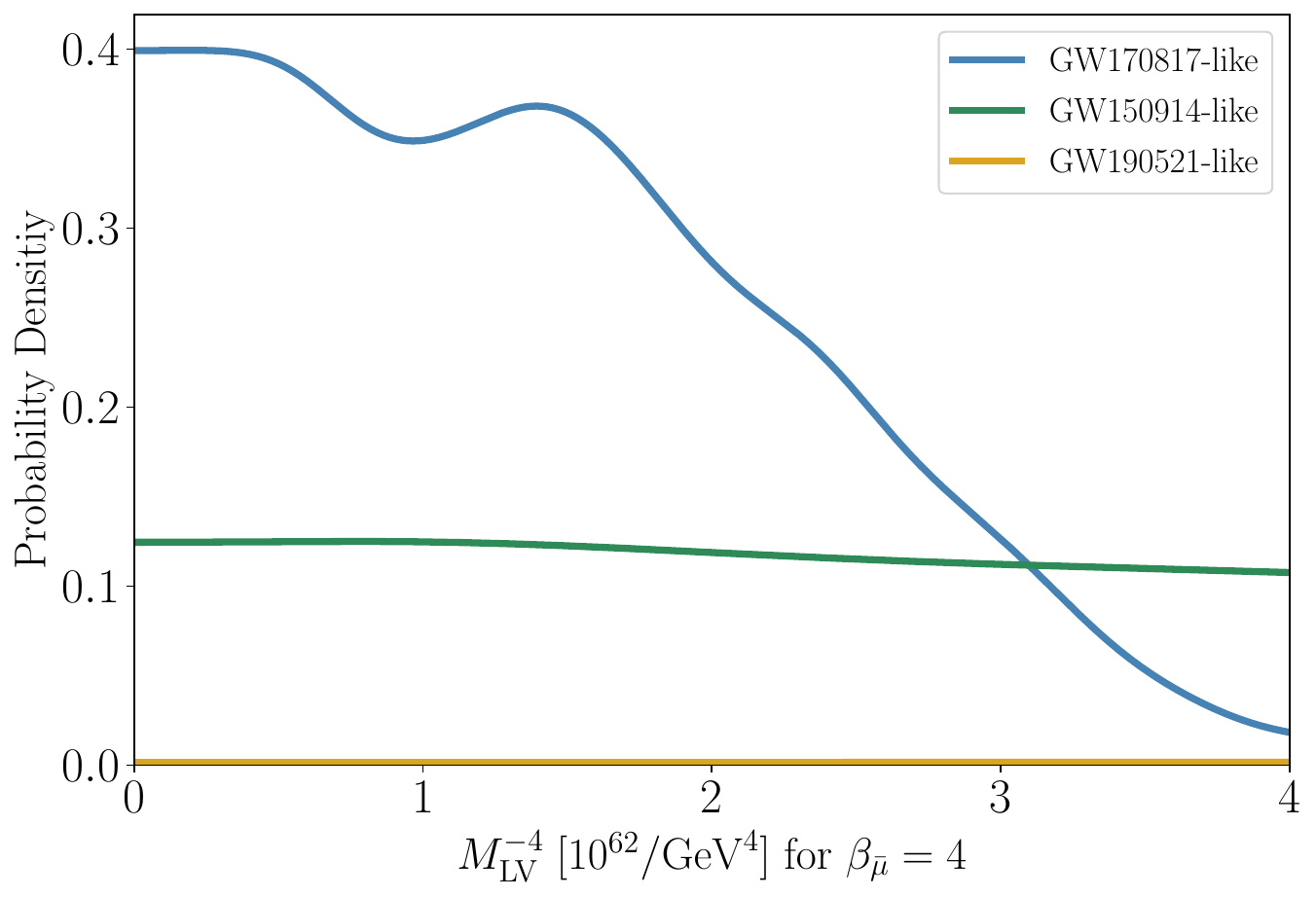}
\caption{The posterior distributions for $M_{\rm PV}^{-\beta_{\nu}}$ with $\beta_{\nu}=1$, $M_{\rm PV}^{-\beta_{\mu}}$ with $\beta_{\mu}=-1, 1, 3$, $M_{\rm LV}^{-\beta_{\bar \nu}}$ with $\beta_{\bar{\nu}}=2$, and $M_{\rm LV}^{-\beta_{\bar \mu}}$ with $\beta_{\bar{\mu}}=-2, 2, 4$ of three injected GW events GW150914-like, GW170817-like, and GW190521-like given by the ground-based GW detectors ET and CE.
\label{fig:mpv and mlv from CE and ET}}
\end{figure*}

In this section, we discuss the results of constraints on parity and Lorentz violations from the analysis performed in the above section with the injected GW signals with the modified waveforms (c.f. (\ref{waveforms})). From each analysis, one can infer the posterior distributions of the parameters $A_\nu$, $A_\mu$, $A_{\bar \nu}$, and $A_{\bar \mu}$ by marginalizing over all GR parameters for a single simulated GW signal. With the corresponding posterior distribution of $A_\nu$, $A_\mu$, $A_{\bar \nu}$, and $A_{\bar \mu}$, together with the posterior distribution of redshift $z$, one can obtain the posterior distributions of $M_{\rm PV}^{-\beta_\nu}$, $M_{\rm PV}^{-\beta_\mu}$, $M_{\rm PV}^{-\beta_{\bar \nu}}$, and $M_{\rm PV}^{-\beta_{\bar \mu}}$ by using Eqs.~(\ref{eq:pv_anu_amu}) and (\ref{eq:lv_anu_amu}), respectively. The corresponding posterior distributions for each analysis are presented in Fig.~\ref{fig:mpv and mlv from CE and ET} for the ground-based GW detectors (ET+CE) and Fig.~\ref{fig:mpv and mlv from LTT} for the space-based GW detectors. For comparison, we also present the posterior distributions of $M_{\rm PV}^{-\beta_\nu}$, $M_{\rm PV}^{-\beta_\mu}$, $M_{\rm PV}^{-\beta_{\bar \nu}}$, and $M_{\rm PV}^{-\beta_{\bar \mu}}$ from the analysis on the corresponding GW events from LVK in Fig.~\ref{fig:mpv and mlv from LVK} by using the results in Refs.~\cite{Zhu:2023rrx, Wang:2025fhw, Wang:2021gqm}. Finally, we straightly translated these posterior distributions into the constraints on $M_{\text{PV}}$ and $M_{\text{LV}}$ for each case and summarized the results in Table \ref{tab:results of ground detectors} for ground-based GW detectors and Table \ref{tab:results of space detectors} for space-based GW detectors, respectively.

In the following subsections, we divided the results into two parts for analysis: constraints from ground-based GW detectors and constraints from space-based GW detectors.

\subsection{Constraints from the third-generation ground-based GW detectors}

\begin{figure*}
\centering
\includegraphics[width=4.4cm]{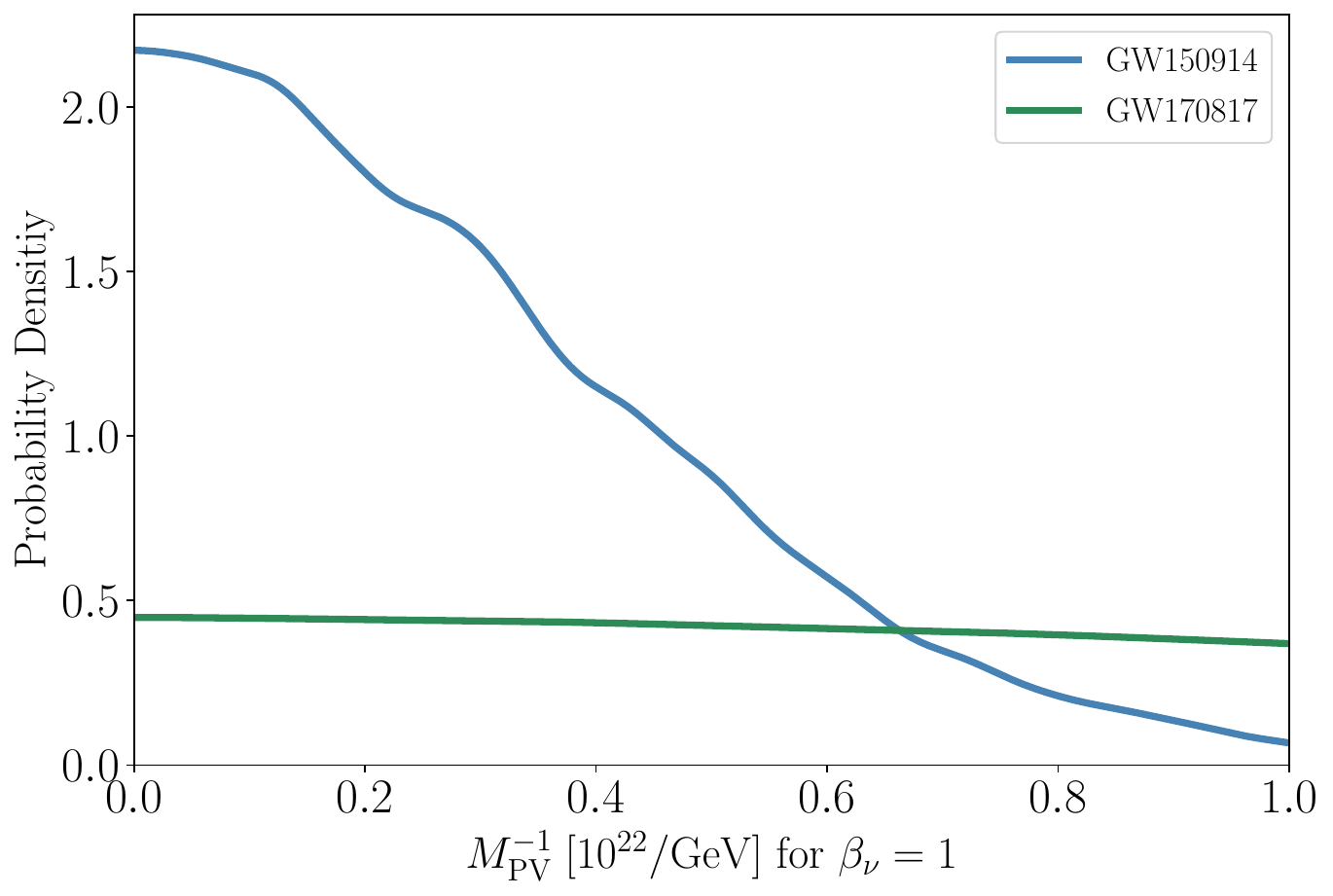}
\includegraphics[width=4.4cm]{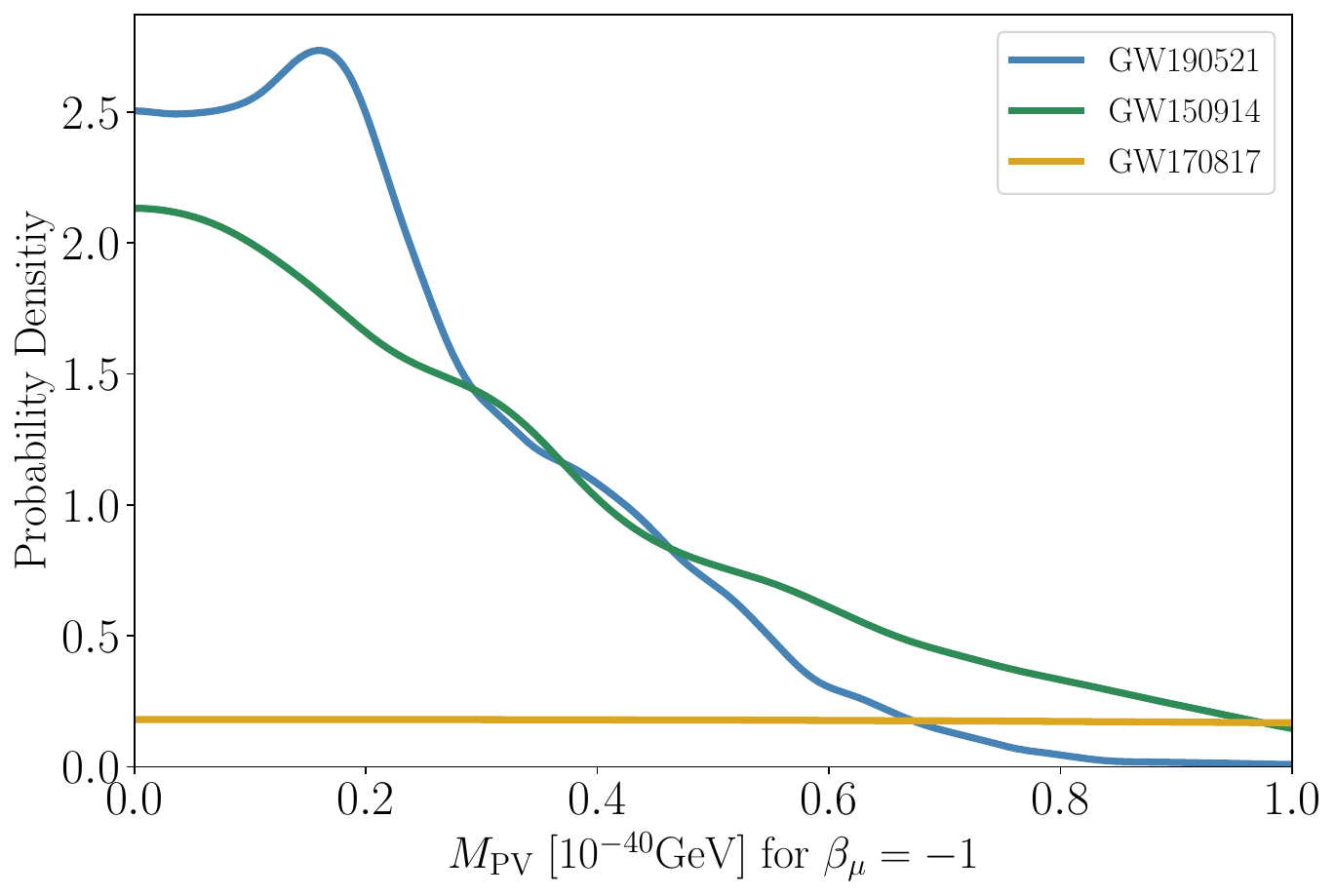}
\includegraphics[width=4.4cm]{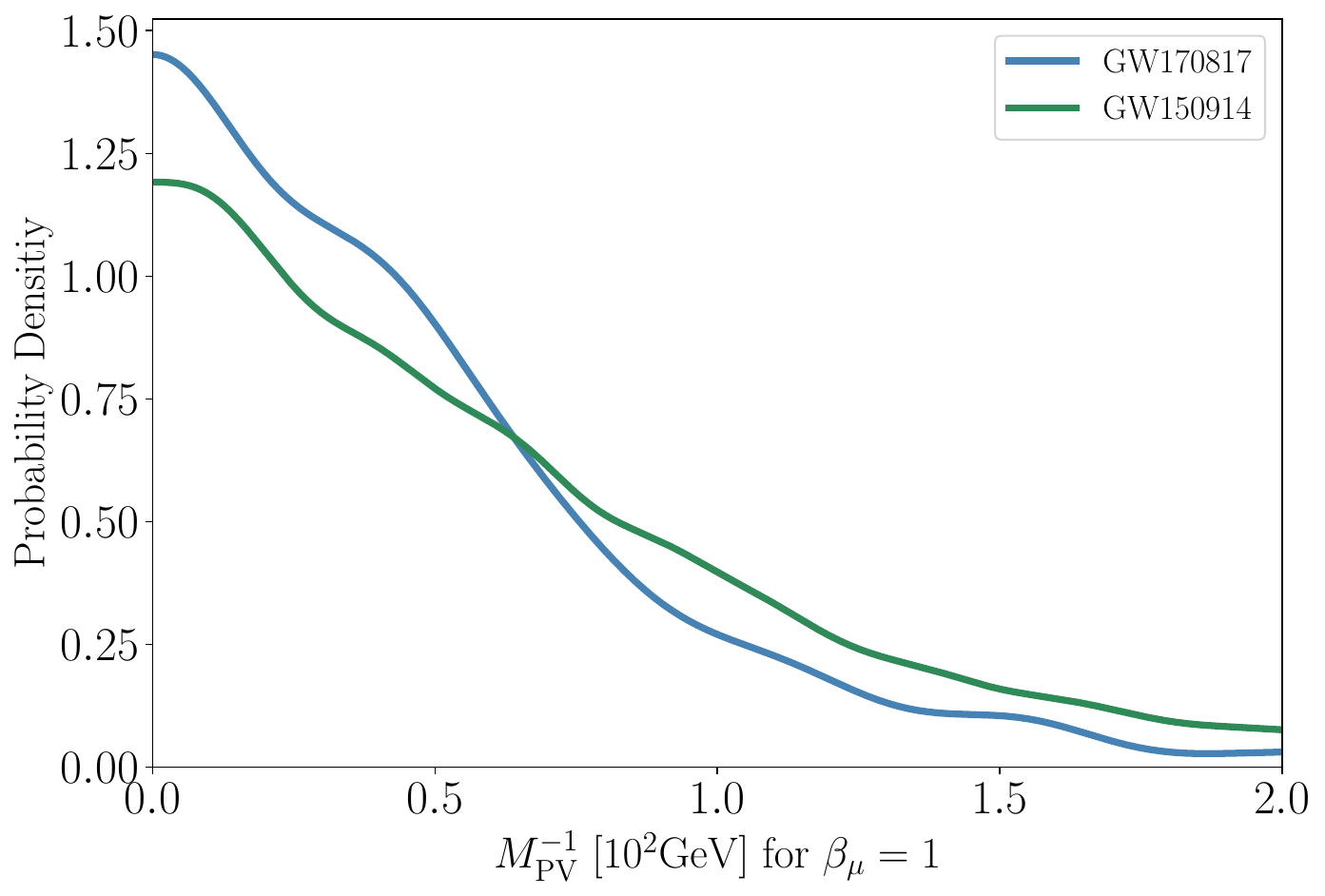}
\includegraphics[width=4.4cm]{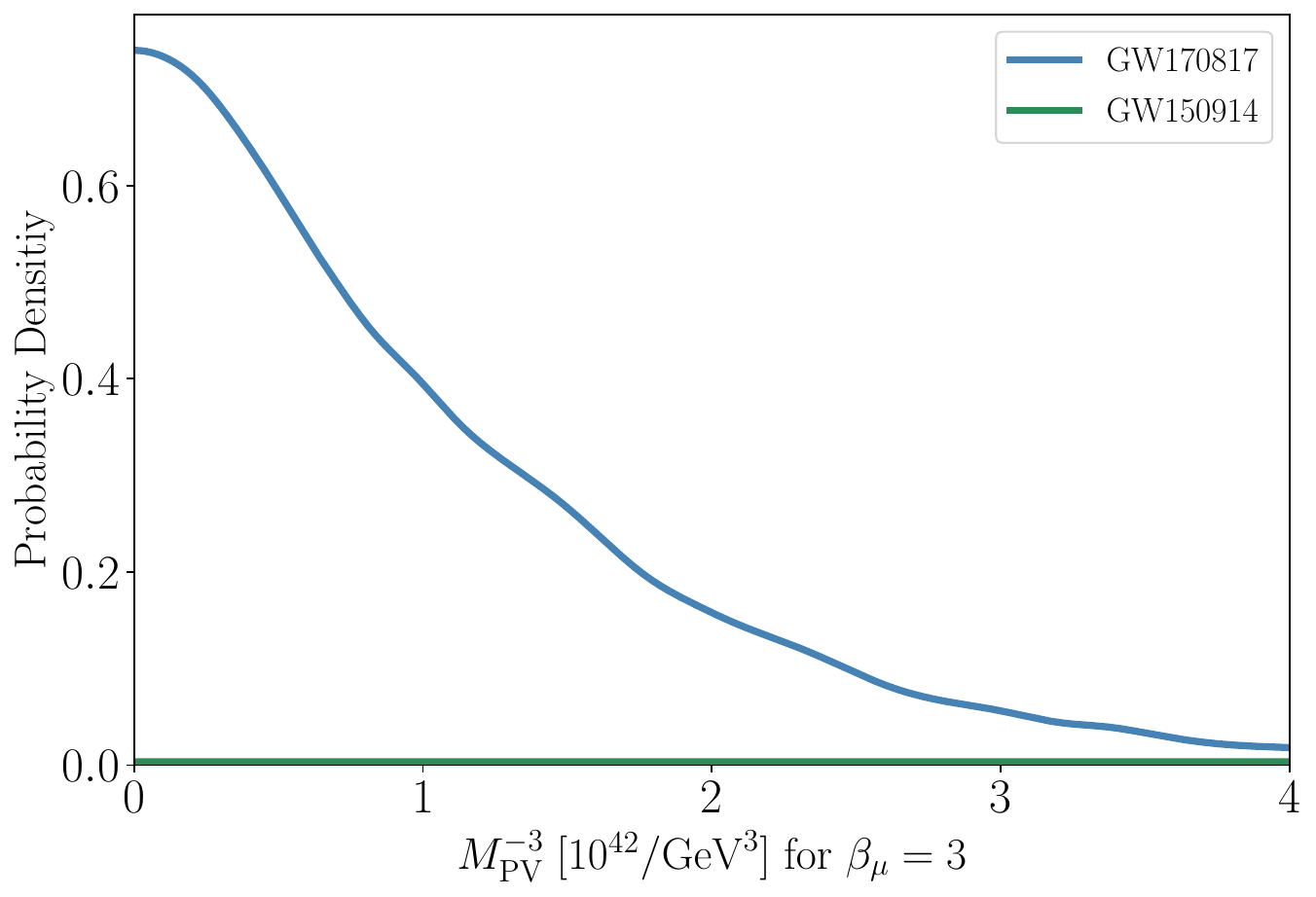}
\includegraphics[width=4.4cm]{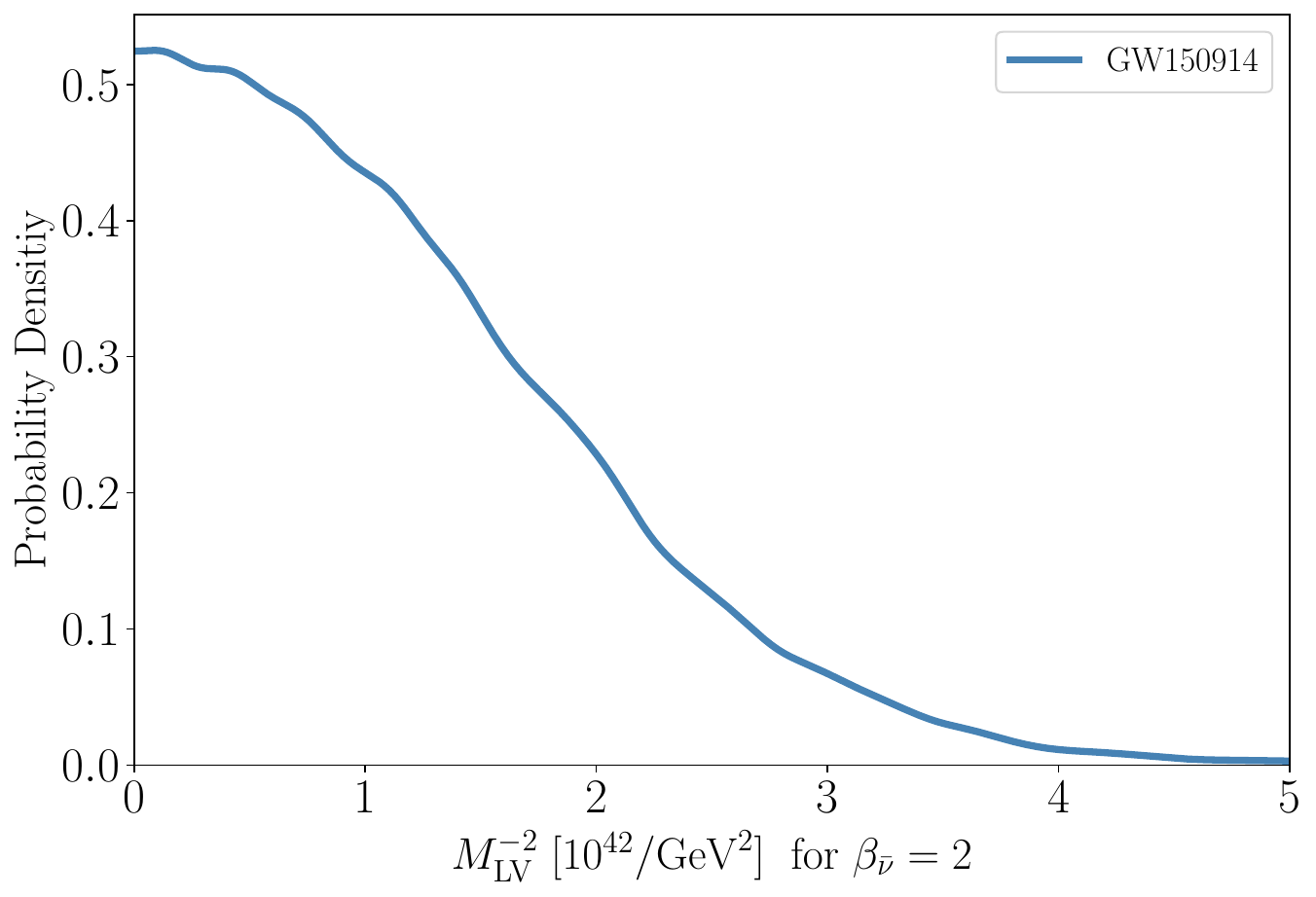}
\includegraphics[width=4.4cm]{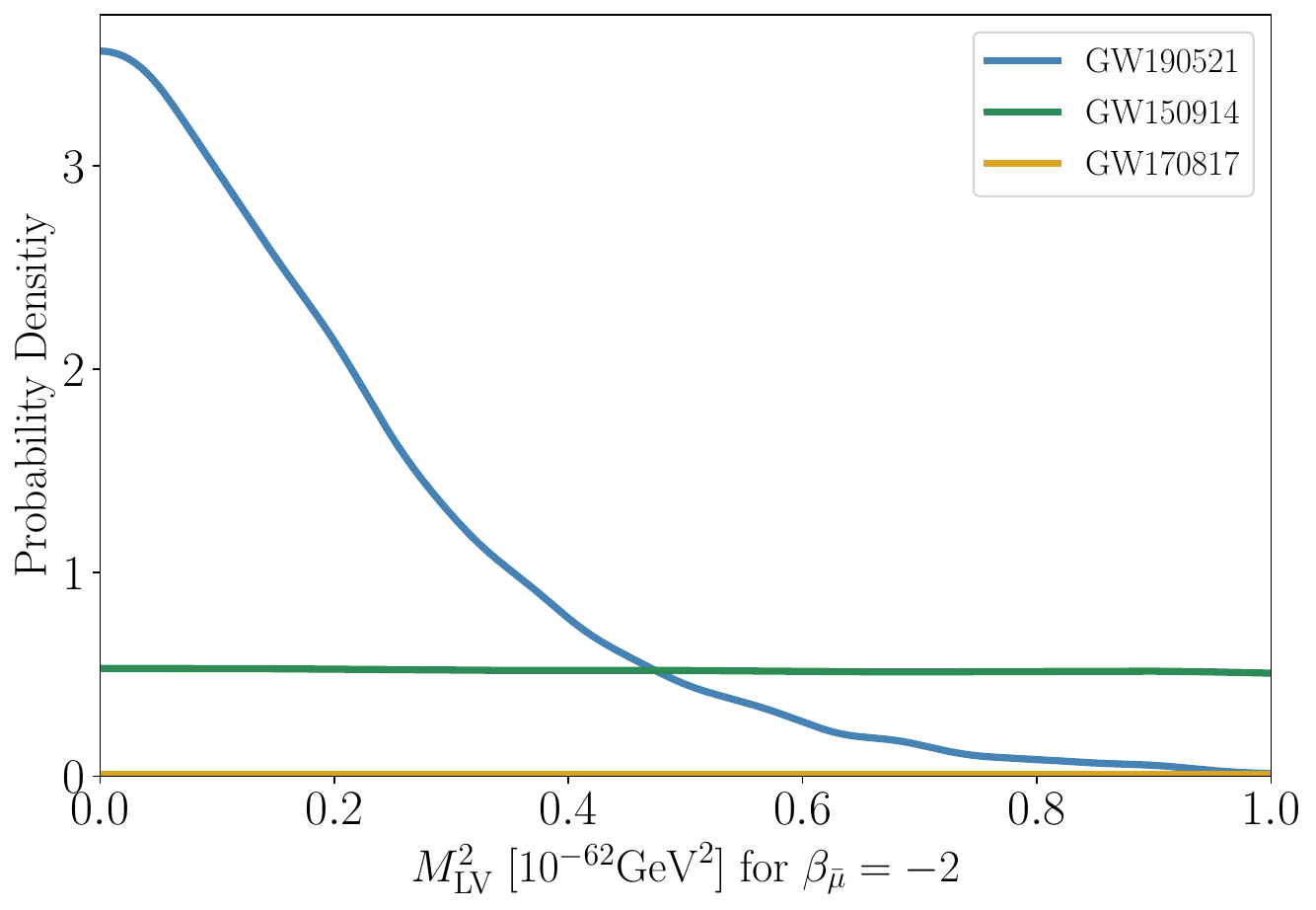}
\includegraphics[width=4.4cm]{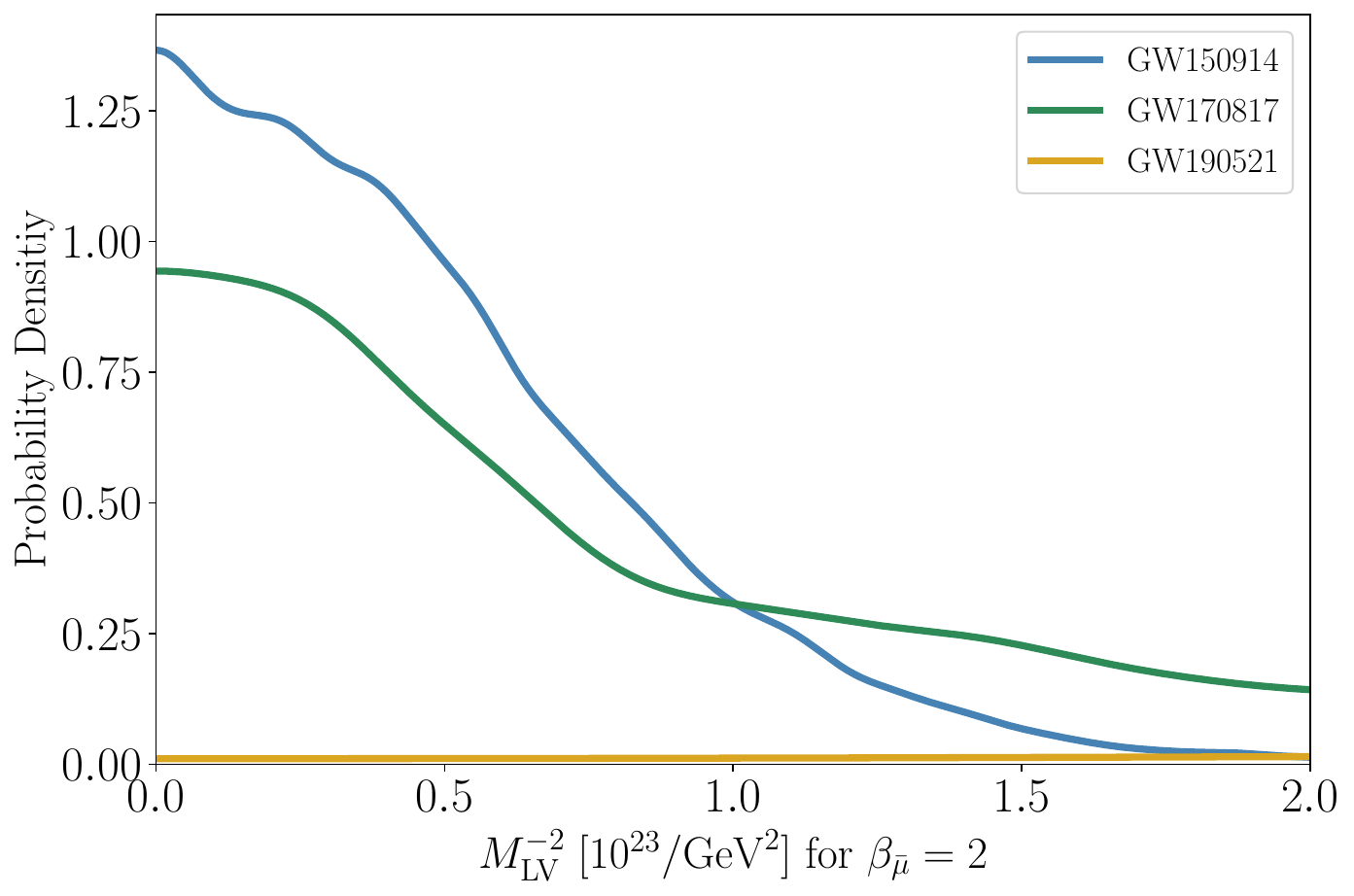}
\includegraphics[width=4.4cm]{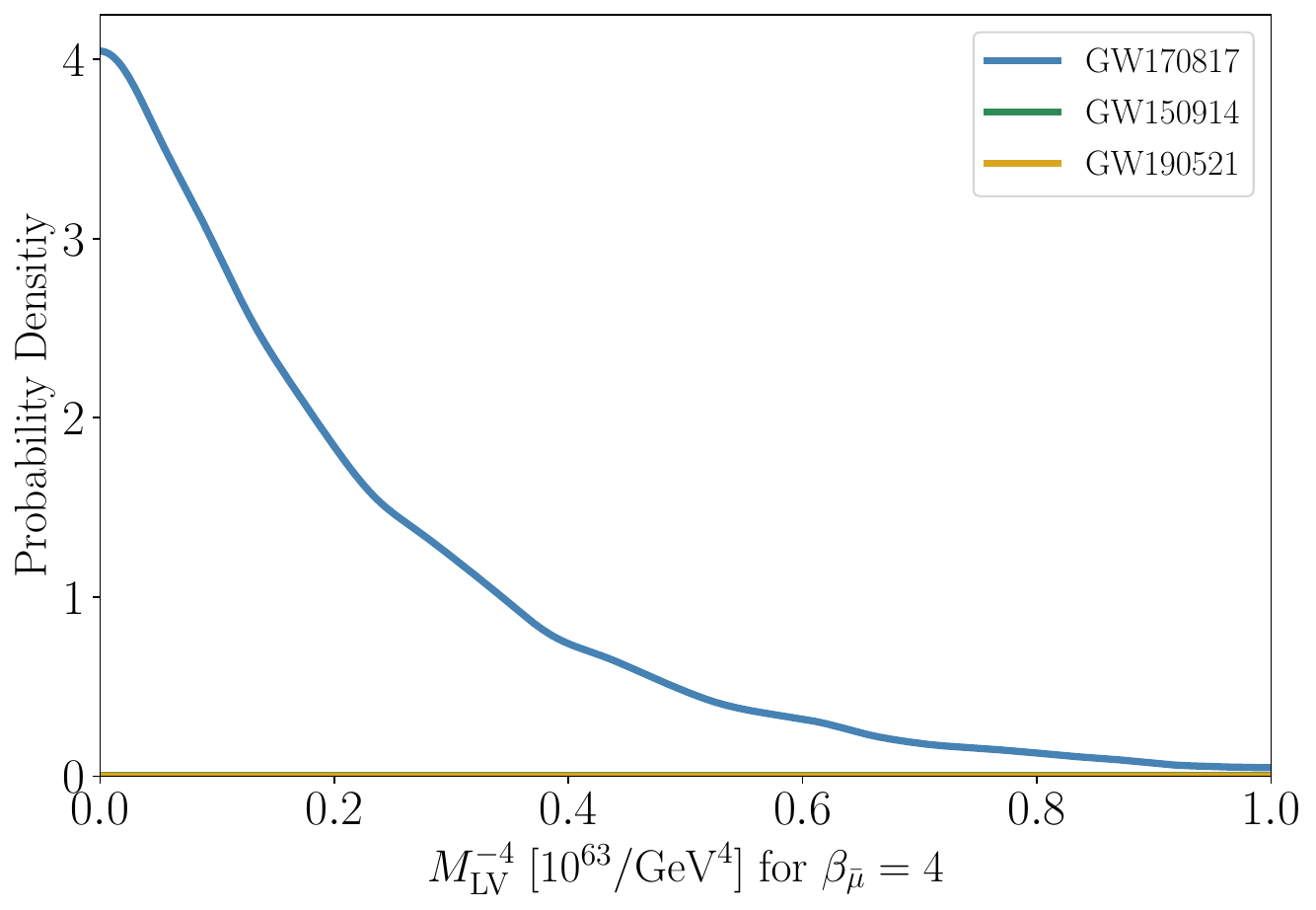}
\caption{The posterior distributions for $M_{\rm PV}^{-\beta_{\nu}}$ with $\beta_{\nu}=1$, $M_{\rm PV}^{-\beta_{\mu}}$ with $\beta_{\mu}=-1, 1, 3$, $M_{\rm LV}^{-\beta_{\bar \nu}}$ with $\beta_{\bar{\nu}}=2$, and $M_{\rm LV}^{-\beta_{\bar \mu}}$ with $\beta_{\bar{\mu}}=-2, 2, 4$ from the analysis with the GW events GW150914, GW170817, and GW190521 from the GWTC-3 of LVK \cite{Zhu:2023rrx, Wang:2025fhw, Wang:2021gqm}. 
\label{fig:mpv and mlv from LVK}}
\end{figure*}

Firstly, let us analyze the constraints on parity and Lorentz violation from the ground-based GW detectors. To show the capabilities of the third-generation of ground-based GW detectors ET+CE for constraining the parity and Lorentz violations, we calculate the $M_{\rm PV}$ and $M_{\rm LV}$ from the analysis of the three simulated GW events, the GW150914-like, GW170817-like, and GW190521-like events. These results are presented in Fig.~\ref{fig:mpv and mlv from CE and ET}, and Table.~\ref{tab:results of ground detectors}. We also present the previous results \cite{Zhu:2023rrx, Wang:2025fhw, Wang:2021gqm} from analysis with the GW data of LVK for comparison, which are presented in Fig.~\ref{fig:mpv and mlv from LVK} and Table.~\ref{tab:results of ground detectors}. 

For parity- and Lorentz-violating effects in the modified waveforms, different values of $\beta_{\nu}$, $\beta_{\mu}$, $\beta_{\bar \nu}$, and $\beta_{\bar \mu}$ correspond to different frequency-dependence of the amplitude and phase corrections. From the results presented in Fig.~\ref{fig:mpv and mlv from CE and ET} and Table.~\ref{tab:results of ground detectors}, one observes that the light event, like GW170817-like event, gives the best constraint on $M_{\rm PV}$ and $M_{\rm LV}$ for large and positive values of $\beta_{\mu}$ and $\beta_{\bar \mu}$, where the phase corrections are largest at high frequencies. The massive events, GW190521-like and GW150914-like events, give roughly better constraints for smaller values of $\beta_{\mu}$ and $\beta_{\bar \mu}$. For the cases with amplitude modulation in the modified waveform ($\beta_{\nu}=1$ and $\beta_{\bar \nu}=2$), the massive events, GW190521-like and GW150914-like events, give best and comparable constraints.

For comparison with previous results, we present the previous results from analysis with the corresponding events, GW150914, GW170817, and GW190521 given from the GW data of LVK in Fig.~\ref{fig:mpv and mlv from LVK} and Table.~\ref{tab:results of ground detectors}. Table.~\ref{tab:results of ground detectors} also shows the combined constraints on $M_{\rm PV}$ and $M_{\rm LV}$ by combining the overall about 90 GW events in GWTC-3 \footnote{For each case, a few events in GWTC-3 has been excluded in the combined analysis, as described in Table.~II of Ref.~\cite{Wang:2025fhw} for $\beta_{\bar \mu}=-2$, in Sec.III.A of Ref.~\cite{Zhu:2023rrx} for $\beta_{\mu}=1$, and in Table.~III of Ref.~\cite{Wang:2021gqm} for other cases.}, see refs. \cite{Zhu:2023rrx}. For individual event, the constraints on $M_{\rm PV}$ for $\beta_{\nu}=1$ and $\beta_{\mu}=-1, 1, 3$ and on $M_{\rm LV}$ for $\beta_{\bar \nu}=2$ and $\beta_{\bar \mu}=-2,2,4$ roughly improves those given in Ref. \cite{Zhu:2023rrx} by 1 - 3 orders of magnitude.

For parity violations, in the cases of the velocity birefringence effect, the most significant improvement is from the case $\beta_{\mu} = -1$, which improves 263 times at least by the joint of ET and CE compared to those given in \cite{Zhu:2023rrx}. For the cases $\beta_{\mu}=1$, the joint ET and CE can give a larger energy scale than the results from LVK by a factor of 20 at least. This result is also consistent with that obtained in Ref.~\cite{Califano:2023aji} from different network configurations of CE and ET. The improvement of case $\beta_{\mu}=3$ is not significant. In the case of amplitude birefringence $\beta_{\nu}=1$, the improvement from ET+CE is more than 20 times. For the case of amplitude birefringence with $\beta_\nu = 1$, our constraint on $M_{\rm PV}$ is slightly less stringent but remains comparable to the result reported in Ref.~\cite{Califano:2023aji}.

For Lorentz violations, compared to the $M_{\rm LV}$ from LVK, ET+CE can improve the $M_{\rm LV}$ about 20 times in binary neutron star systems and about 18 times in binary black hole systems. The minimum improvement of case $\beta_{\bar{\mu}}=2$ and $\beta_{\bar{\mu}}=4$, which are also the effects from the theories with modified dispersion relation are 10.7 and 1.1, respectively.

\begin{table*}
\caption{\label{tab:results of ground detectors}%
Results from the Bayesian analysis of the parity- and Lorentz-violating waveforms on GW events in LVK and injected signals to be detected by ET+CE. The upper half of the table shows the constraints from a single GW event and the joint constraint from more than 80 GW events in the set of GWTC-3 \cite{Zhu:2023rrx, Wang:2021gqm}. Some previous works \cite{Zhao:2022pun, Wang:2020pgu} concentrate on the case $\beta_{\mu}=1$, and we reveal the results from Ref.~\cite{Wang:2021gqm} for comparison. The lower half of the table shows the constraints from three simulated GW signals, GW150914-like, GW170817-like, and GW190521-like, with the third-generation GW detectors CE and ET. The table shows 90\%-credible upper bounds on $M_{\rm PV}$ for $\beta_{\mu}=-1$ (for velocity birefringence) and $M_{\rm LV}$ for $\beta_{\bar{\mu}}=-2$. In the other cases, the results show the lower bounds on $M_{\rm PV}$ and $M_{\rm LV}$.}
\begin{ruledtabular}
{
\begin{tabular}{ccccccccc}
 & \multicolumn{4}{c}{$M_{\rm PV}$ [GeV]} &\multicolumn{4}{c}{$M_{\rm LV}$ [GeV]} \\
\cline{2-5}  \cline{6-9}
  & $\beta_\nu=1$ & $\beta_\mu=-1$ & $\beta_\mu=1$ & $\beta_\mu=3$ & $\beta_{\bar \nu}=2$ & $\beta_{\bar \mu}=-2$ &$\beta_{\bar \mu}=2$ & $\beta_{\bar \mu}=4$    \\
  & $[10^{-23}]$ & $[10^{-43}]$ & $[10^{-3}]$ & $[10^{-15}]$ & $[10^{-22}]$ & $[10^{-33}]$ &$[10^{-12}]$ & $[10^{-17}]$    \\
\colrule
GW170817 LVK    & $3.1$   & $13000$    & $8.4$ 
\cite{Wang:2021gqm}
& $7.5$   &  ---     &  $1100$ \cite{Wang:2025fhw} & $2.1$     & $21.1$  \\
GW150914 LVK    & $16$    & $750$    & $7.1$ \cite{Wang:2021gqm} 
&$1.9$  & $6.4$    &  $150$ \cite{Wang:2025fhw} & $3.1$     & $4.9$   \\
GW190521 LVK    & ---    & $480$   & --- 
& ---   & ---    & $68$ \cite{Wang:2025fhw} & ---  & ---  \\
GWTC-3 combined & $40$    & $80$      & $50$ \cite{Wang:2021gqm}  
& $12$    & $14$     & $83.1$ \cite{Wang:2025fhw}
& $12$    & $34$ \\
\hline
GW170817 ET+CE & $65$   & $2.0$   & $390$  & $11$    &  $17$   & $51.5$  & $25.2$  & $24$ \\
GW150914 ET+CE & $364$  & $0.36$    & $1720$  & $8.9$    &  $32.8$  & $11.8$ & $33.4$ & $17.7$ \\
GW190521 ET+CE & $375$  & $1.82$   & $547$  & $4.5$   &  $31.1$  & $3.7$  & $15.2 $ & $8.3$ \\
\end{tabular}}
\end{ruledtabular}
\end{table*}

\subsection{Constrains from the space-based GW detectors}

In this subsection, we analyze the constraints imposed by space-based GW detectors on parity and Lorentz violations. Specifically, we focus on three planned space-based GW detectors, LISA, Taiji, and TianQin, which are expected to commence operations in the 2030s. In the absence of direct observations, we consider three hypothetical equal-mass binary merger events: event 1 ($M=2 \times 10^4 \, M_{\odot}, \, z=1$), event 2 ($M=2 \times 10^5 \, M_{\odot}, \, z=5$), and event 3 ($M=2 \times 10^6 \, M_{\odot}, \, z=10$). These events are used to evaluate the ability of the GW detectors to constrain parity and Lorentz violations. For our analysis, we consider two GW detector network combinations: LISA+Taiji and LISA+TianQin, referencing the discussion in Ref.~\cite{Wang:2021srv}. Additionally, a more optimistic scenario involving the combination of all three GW detectors has been proposed in Ref.~\cite{Jin:2023sfc}. Overall, we find that the constraints derived from the LISA+Taiji and LISA+TianQin combinations are comparable.

For cases with positive values of $\beta_{\nu}$, $\beta_\mu$, $\beta_{\bar \nu}$, and $\beta_{\bar \mu}$, the constraints on $M_{\rm PV}$ and $M_{\rm LV}$ are generally weaker than those from the ground-based GW detectors by 3 -4 orders of magnitude. This is easy to understand since the amplitude and phase corrections for these cases are proportional to $\sim f^{\beta_{\nu}}, f^{\beta_{\bar \nu}}, f^{\beta_{\mu}+1}, f^{\beta_{\bar \mu}+1}$. They are more relevant for high frequencies of GWs. For these cases, ground-based GW detectors show a clear advantage. For $\beta_{\mu}=-1$ case, the constraints on $M_{\rm PV}$ given by the space-based GW detector are weaker than those from ET+CE but stronger than those from LVK data. Constraints improve with increasing mass and redshift of the compact binary system. In the case of $\beta_{\bar{\mu}}=-2$, space-based GW detectors provide constraints three orders of magnitude tighter than LVK results \cite{LIGOScientific:2021sio}. This aligns with expectations, as lower-frequency GWs yield a stricter upper bound on $M_{\rm LV}$ for this case. This constraint can also transform to an upper bound on the graviton mass, which is consistent with the forecasted results from the Fisher matrix \cite{Mirshekari:2011yq, Will:1997bb}.

\begin{figure*}
\centering
\includegraphics[width=4.4cm]{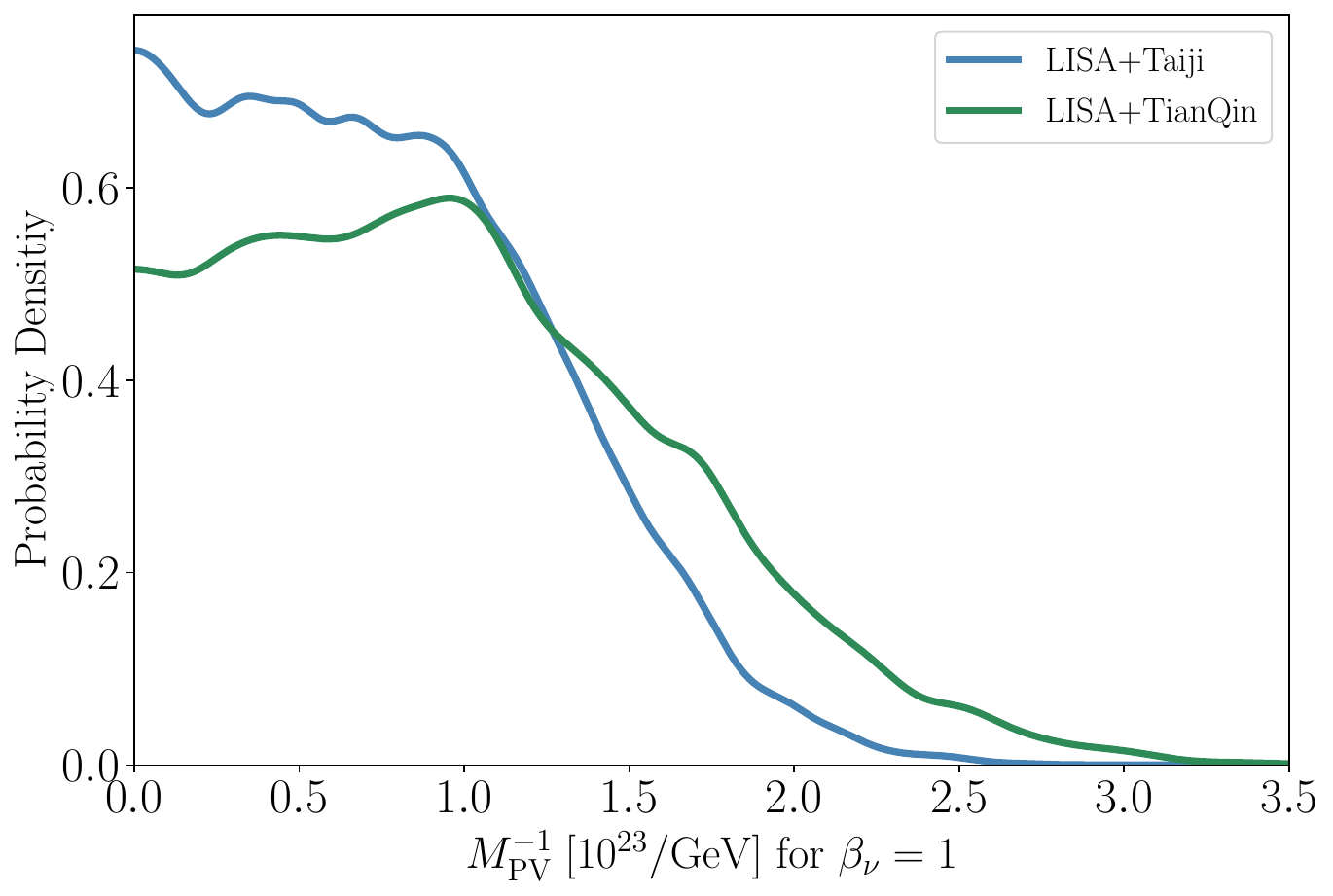}
\includegraphics[width=4.4cm]{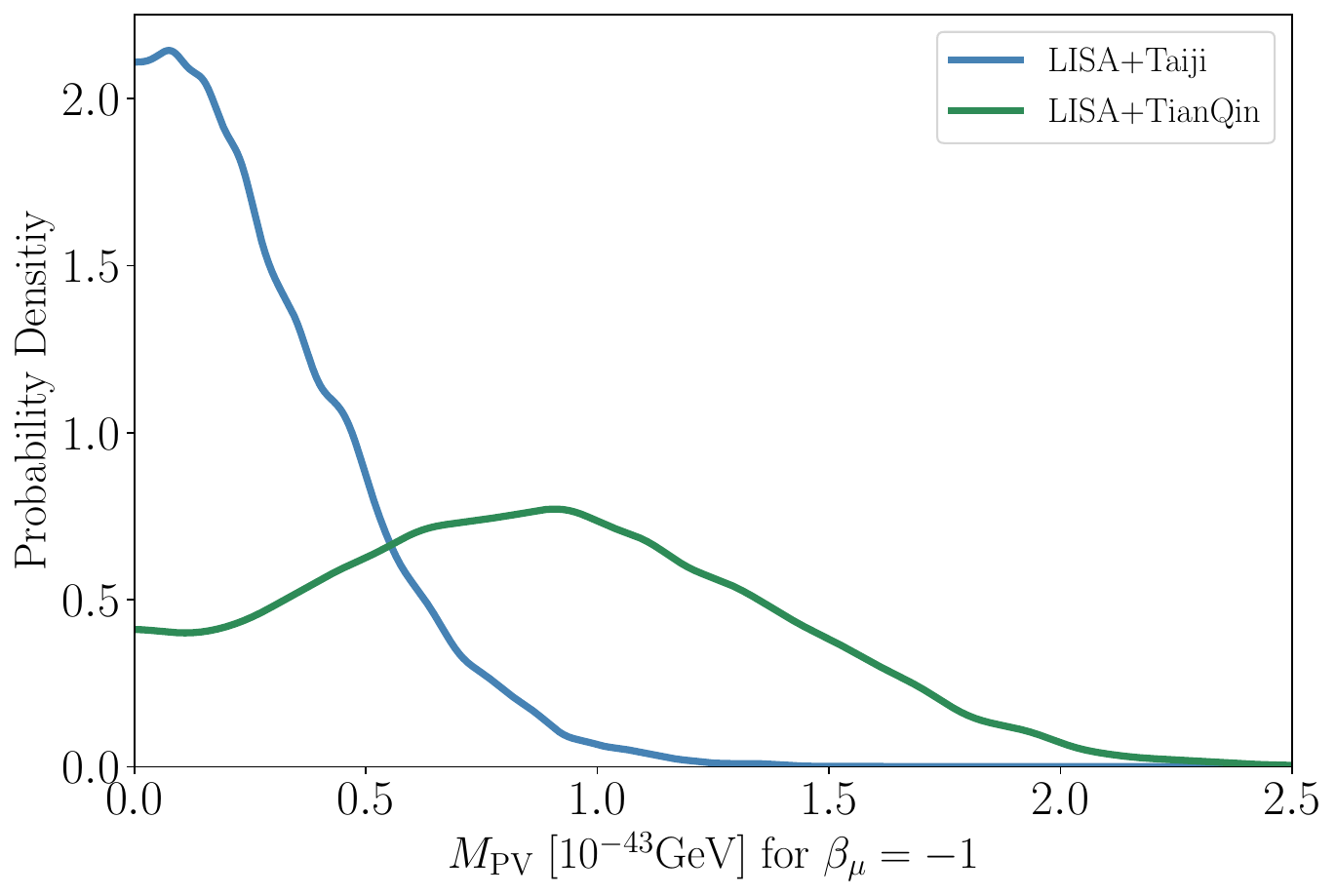}
\includegraphics[width=4.4cm]{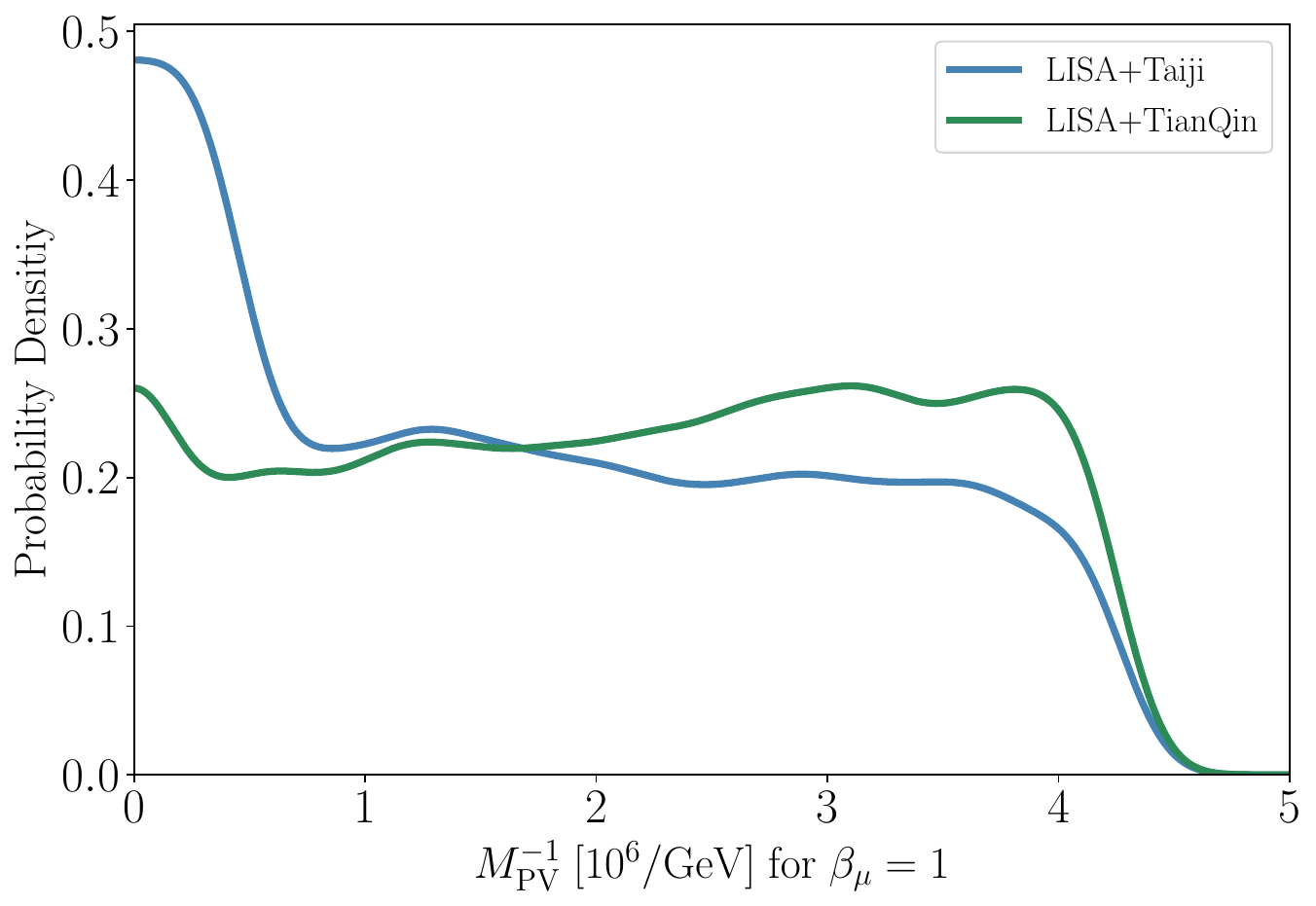}
\includegraphics[width=4.4cm]{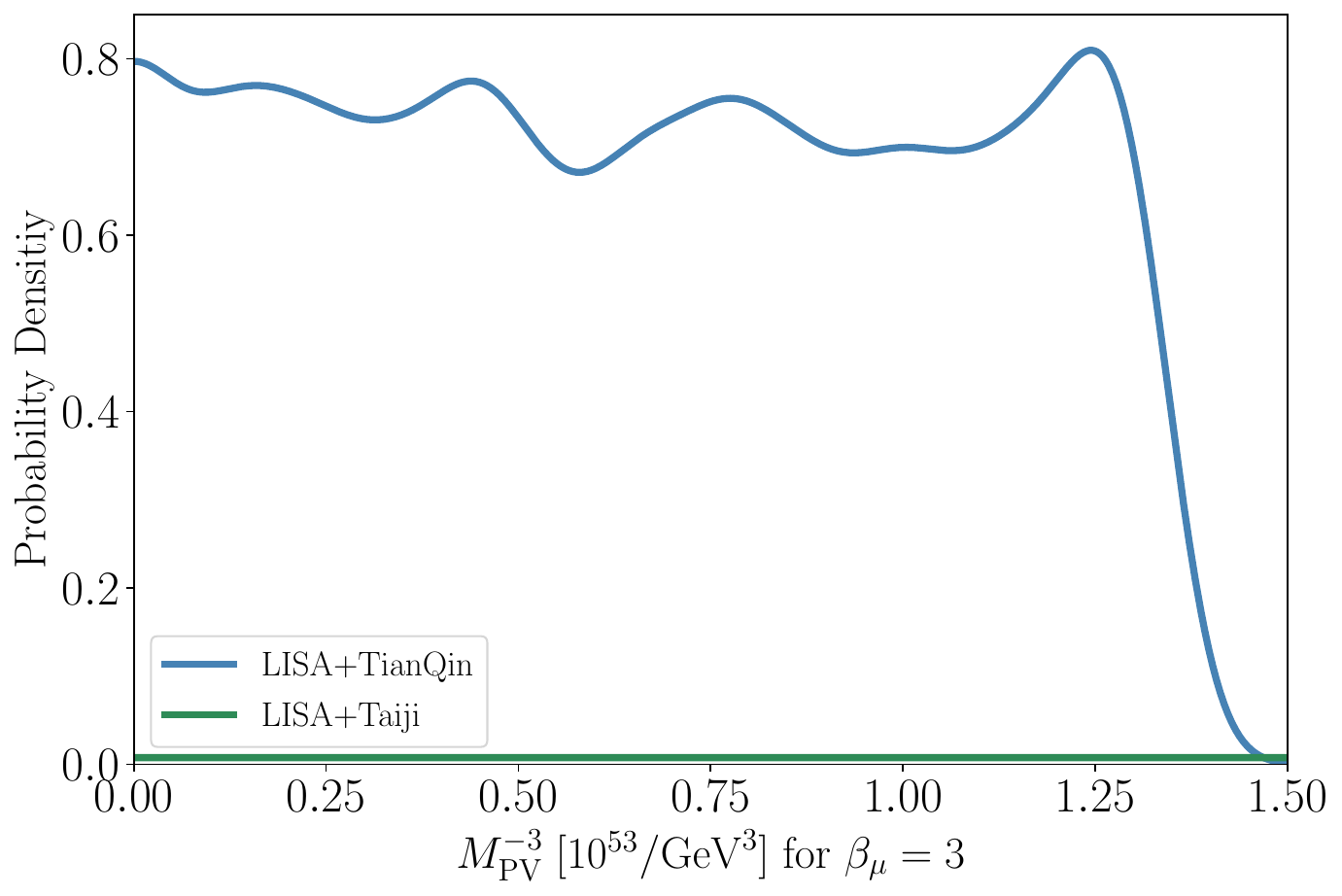}
\includegraphics[width=4.4cm]{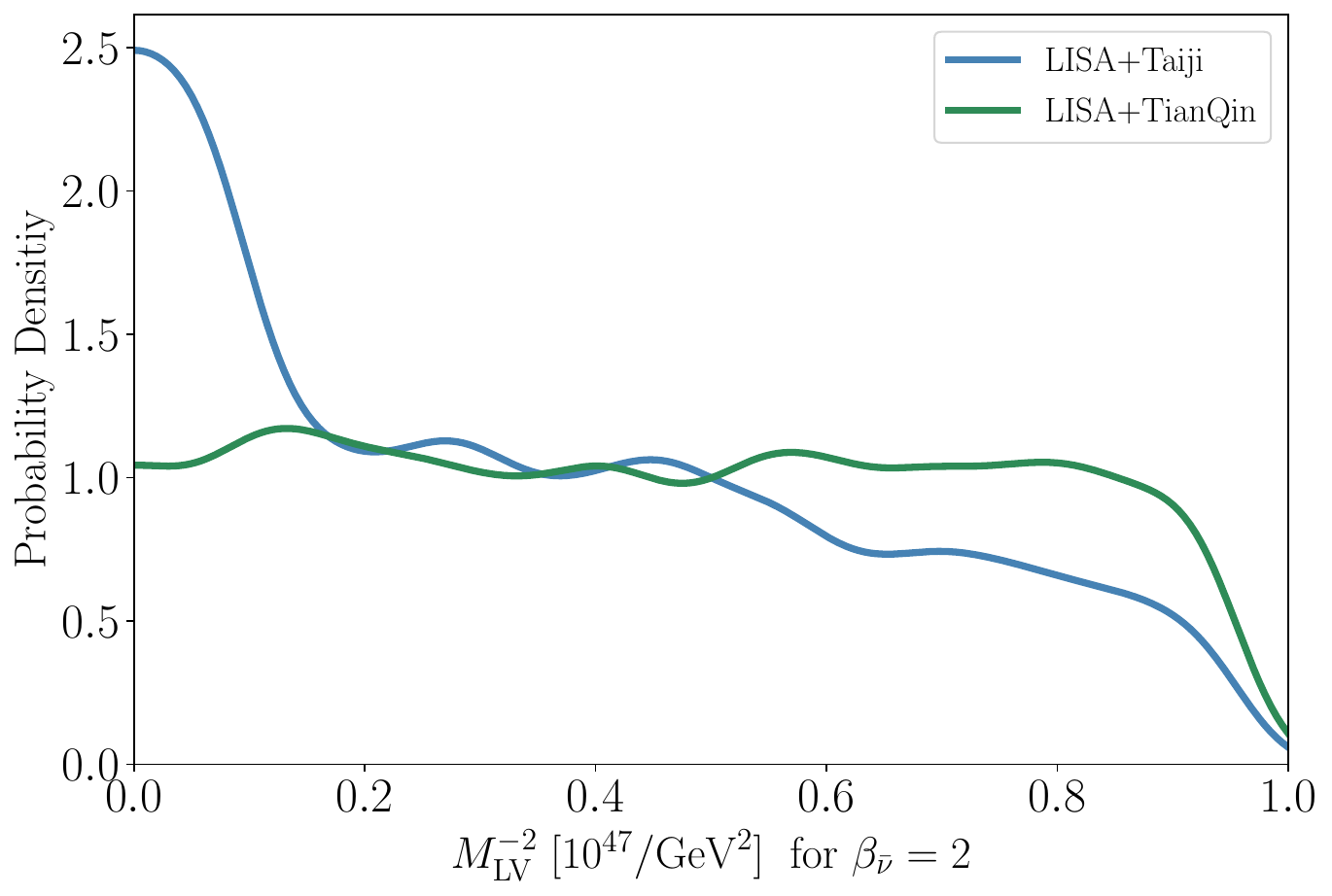}
\includegraphics[width=4.4cm]{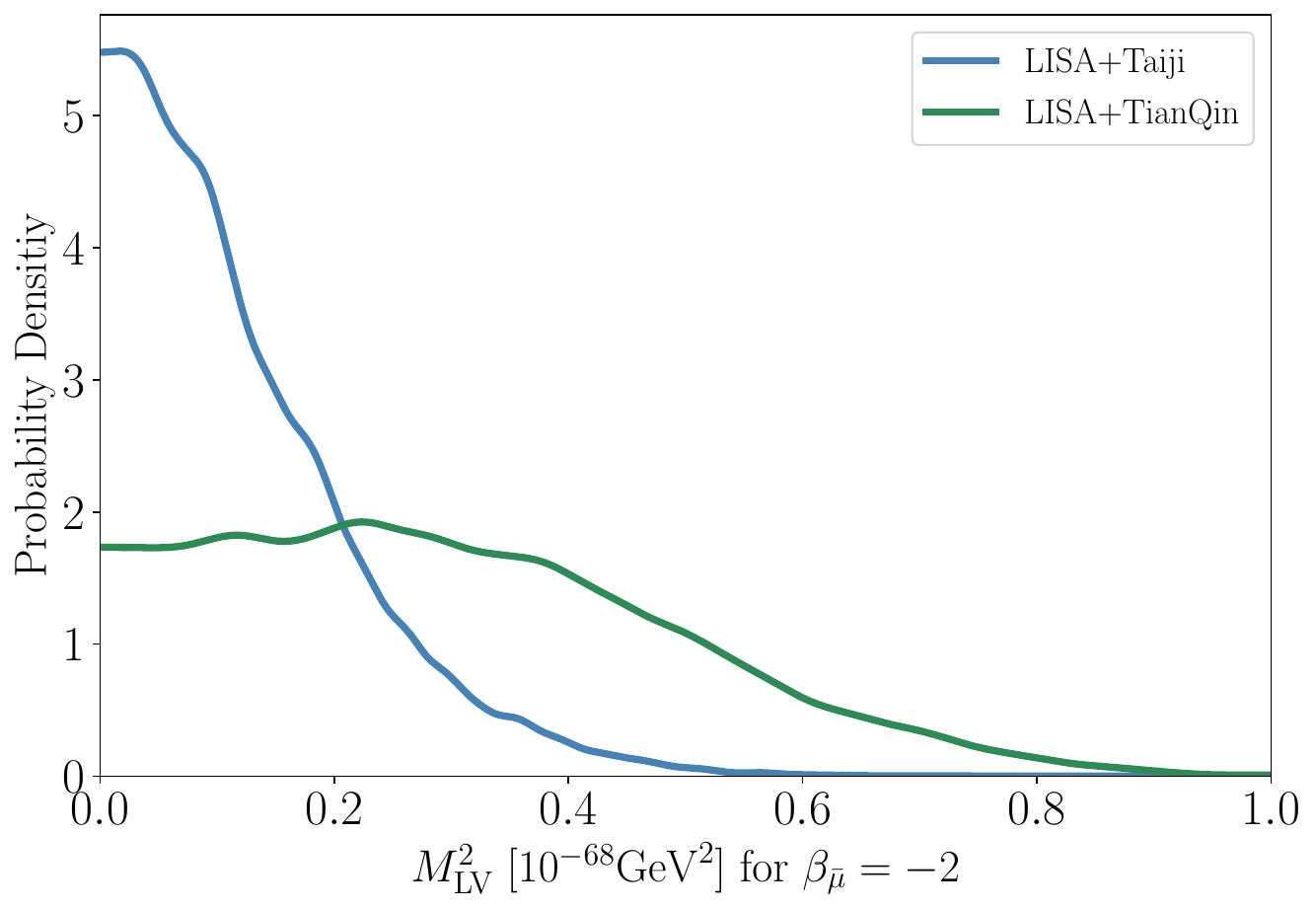}
\includegraphics[width=4.4cm]{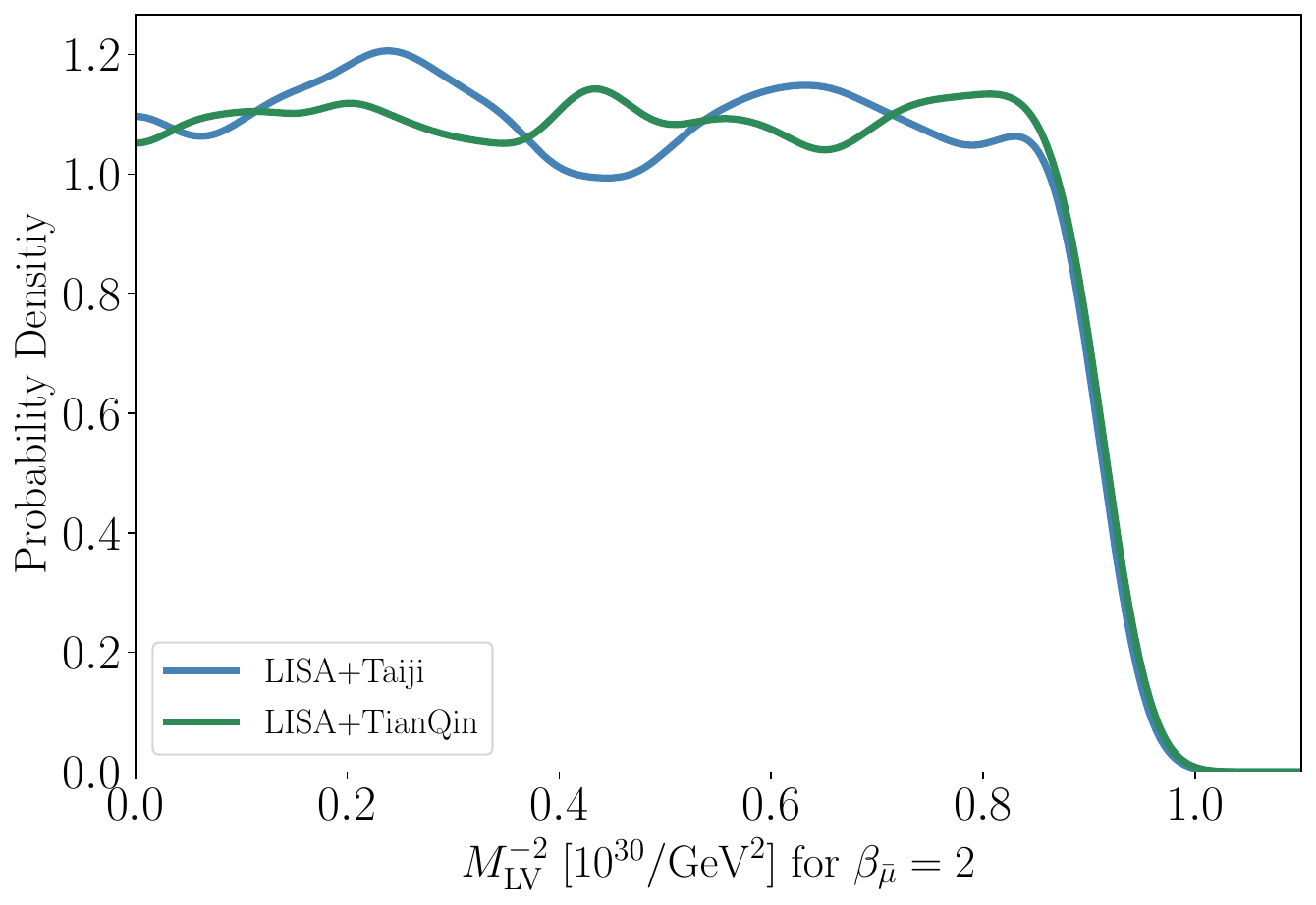}
\includegraphics[width=4.4cm]{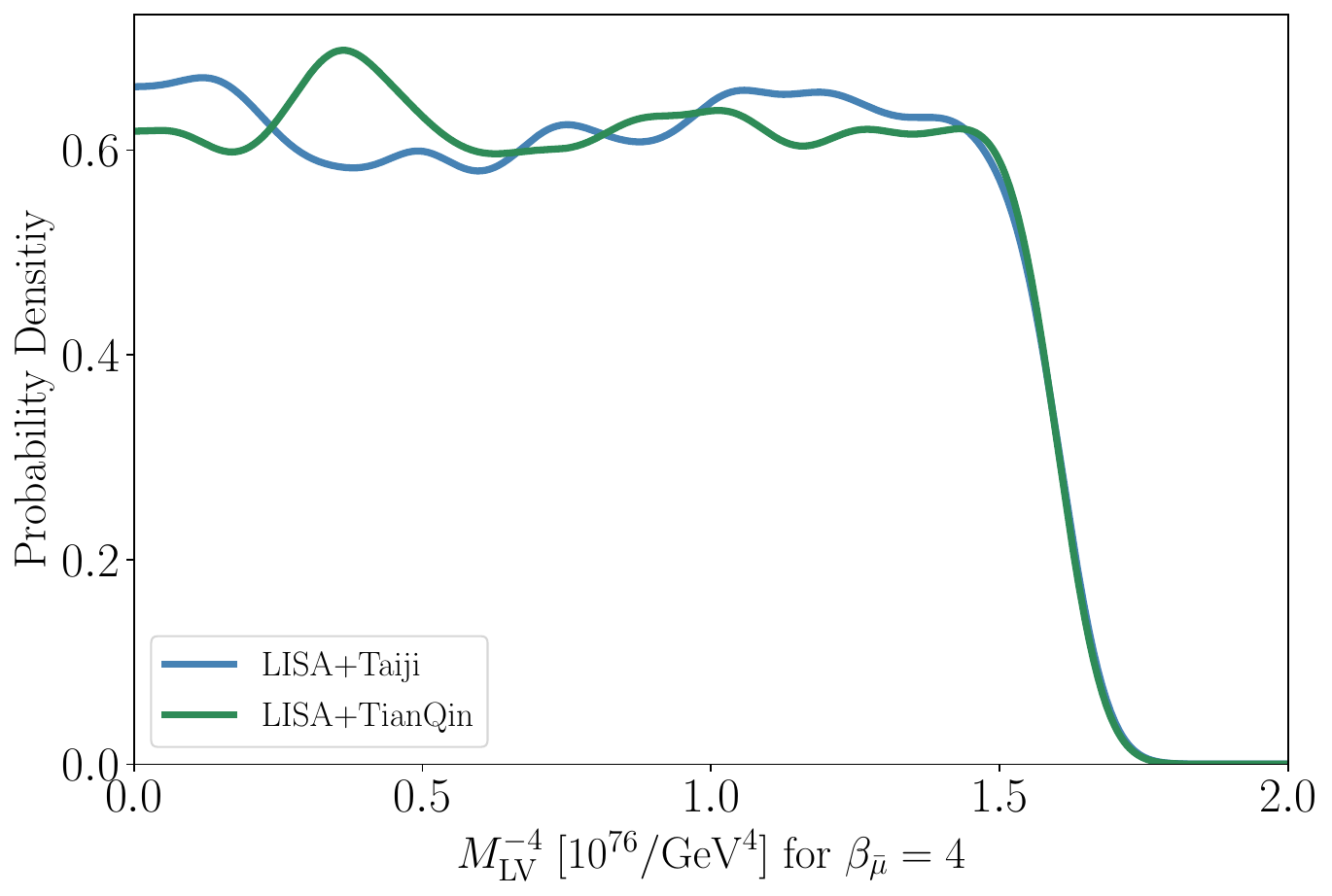}
\caption{The posterior distributions for $M_{\rm PV}^{-\beta_{\nu}}$ with $\beta_{\nu}=1$, $M_{\rm PV}^{-\beta_{\mu}}$ with $\beta_{\mu}=-1, 1, 3$, $M_{\rm LV}^{-\beta_{\bar \nu}}$ with $\beta_{\bar{\nu}}=2$, and $M_{\rm LV}^{-\beta_{\bar \mu}}$ with $\beta_{\bar{\mu}}=-2, 2, 4$ of the three simulated events for the space-based GW detectors.
\label{fig:mpv and mlv from LTT}}
\end{figure*}

\begin{table*}
\caption{\label{tab:results of space detectors}%
Results from the Bayesian analysis of the parity- and Lorentz-violating waveforms on injected GW signals to be detected by LISA+Taiji and LISA+TianQin. The upper half of the table shows the results of the constraint from a single GW event by a combination of LISA and Taiji. The lower half of the table shows the constraints from the combination of LISA and TianQin. The table shows 90\%-credible upper bounds on $M_{\rm PV}$ for $\beta_{\mu}=-1$ (for velocity birefringence) and $M_{\rm LV}$ for $\beta_{\bar{\mu}}=-2$. In the other cases, the results show the lower bounds on $M_{\rm PV}$ and $M_{\rm LV}$.}
\begin{ruledtabular}
{
\begin{tabular}{ccccccccc}
 & \multicolumn{4}{c}{$M_{\rm PV}$ [GeV]} &\multicolumn{4}{c}{$M_{\rm LV}$ [GeV]} \\
\cline{2-5}  \cline{6-9}
  & $\beta_\nu=1$ & $\beta_\mu=-1$ & $\beta_\mu=1$ & $\beta_\mu=3$ & $\beta_{\bar \nu}=2$ & $\beta_{\bar \mu}=-2$ &$\beta_{\bar \mu}=2$ & $\beta_{\bar \mu}=4$    \\
  & $[10^{-25}]$ & $[10^{-44}]$ & $[10^{-8}]$ & $[10^{-19}]$ & $[10^{-25}]$ & $[10^{-35}]$ &$[10^{-16}]$ & $[10^{-20}]$    \\
\colrule
\hline
event1 LISA+Taiji    & 4.2  & 26    & 4.4   & 1.4   &  8.8   & 8.9  & 2.2 & 2.9   \\
event2 LISA+Taiji    & 37   & 23    & 22    & 3.4   &  20    & 9.0  & 6.8 & 5.3   \\
event3 LISA+Taiji    & 67   & 6.2   & 27    & 4.4   &  36    & 5.1  & 11  & 9.1   \\

\hline
event1 LISA+TianQin  & 7.3  & 34    & 4.9   & 4.3   & 8.7    & 12   & 2.2  & 2.9   \\
event2 LISA+TianQin  & 40   & 21    & 17    & 12    & 24     & 11   & 6.8 & 5.4   \\
event3 LISA+TianQin  & 53   & 16    & 26    & 20    & 34     & 7.6  & 11 & 9.1   \\

\end{tabular}}
\end{ruledtabular}
\end{table*}

\section{CONCLUSION AND DISCUSSION}\label{sec:5}
\renewcommand{\theequation}{4.\arabic{equation}} \setcounter{equation}{0}

In this paper, we investigate the prospects of constraining parity and Lorentz violations in gravity by using the simulated GW signals with the next generation of
ground-based GW detectors (ET and CE) and future space-based GW detectors (LISA, Taiji, and TianQin). We adopt a systematic parametric framework for characterizing possible derivations of GW propagation and modified waveform with parity- and Lorentz-violating effects \cite{Zhu:2023rrx, Zhao:2019xmm}. For ground-based GW detectors, we consider a network consisting of two third-generation GW detectors, ET and CE. For space-based GW detectors, we use two different networks, LISA+Taiji and LISA+TianQin, respectively. 

To forecast the future GW detectors, we use the simulated GW data based on the GW template \texttt{IMRPhenomPv2} and \texttt{IMRPhenomPv2\_NRTidal} of GR. For comparison with previous results from analysis with GW data of LVK \cite{Zhu:2023rrx}, we inject three typical GW events to ET+CE by assuming their parameters as in the events GW150914, GW170817, and GW190521. For space-based GW detectors, we inject three GW signals with parameters given in Table.~\ref{tab:event list}. These signals are then subjected to the specific GW detector networks and we perform Bayesian analysis on the modified waveforms with parity- and Lorentz-violating effects to obtain the prospects for constraining parity ad Lorentz violations in gravity.

We analyzed the modified GW waveforms in Eq.~(\ref{waveforms}) under eight distinct cases, considering parity violation with $\beta_{\nu}=1$ and $\beta_{\mu}=-1, 1, 3$, and Lorentz violation with $\beta_{\bar{\nu}}=2$ and $\beta_{\bar{\mu}}=-2, 2, 4$. The corresponding constraints on the energy scales of parity violation ($M_{\rm PV}$) and Lorentz violation ($M_{\rm LV}$) are summarized in Table~\ref{tab:results of ground detectors} and Table.~\ref{tab:results of space detectors}. The posterior distributions of $M_{\rm PV}^{-\beta_{\nu}}$, $M_{\rm PV}^{-\beta_{\mu}}$, $M_{\rm LV}^{-\beta_{\bar \nu}}$, and $M_{\rm LV}^{-\beta_{\bar \mu}}$ derived from the ground- and space-based GW detectors are shown in Fig.~\ref{fig:mpv and mlv from CE and ET} and Fig.~\ref{fig:mpv and mlv from LTT}, respectively.

Our results demonstrate that the constraints from ET and CE impose significantly tighter bounds on $M_{\rm PV}$ and $M_{\rm LV}$ compared to those obtained with current LVK observations. For cases with positive values of $\beta_{\nu}$, $\beta_{\mu}$, $\beta_{\bar{\nu}}$, and $\beta_{\bar{\mu}}$, ground-based GW detectors (ET+CE) outperform space-based GW detectors due to the stronger relevance of parity- and Lorentz-violating effects at high GW frequencies. However, for $\beta_{\mu} = -1$, space-based GW detectors provide constraints on $M_{\rm PV}$ that are superior to those from LVK and comparable to ET+CE. Similarly, for $\beta_{\bar{\mu}} = -2$, space-based GW detectors exhibit greater sensitivity in constraining $M_{\rm LV}$, with constraints on $M_{\rm PV}$ being approximately three orders of magnitude tighter than those from ground-based GW detectors. Additionally, this scenario allows for bounds on the graviton mass at \( m_g \lesssim 10^{-35} \) GeV, consistent with previous studies. These findings highlight the complementarity of ground- and space-based GW detectors in probing the nature of gravity through GWs. The continued improvement of GW detector sensitivity will be crucial in furthering our understanding of parity and Lorentz symmetry of gravity.

\section*{ACKNOWLEDGMENTS}
Tao Zhu and Bo-Yang Zhang are supported by the National Key Research and Development Program of China under Grant No.2020YFC2201503, the National Natural Science Foundation of China under Grants No.12275238 and No. 11675143, the Zhejiang Provincial Natural Science Foundation of China under Grants No.LR21A050001 and No. LY20A050002, and the Fundamental Research Funds for the Provincial Universities of Zhejiang in China under Grant No. RF-A2019015. 

Jing-Fei Zhang, Xin Zhang, and Bo-Yang Zhang are supported by the National SKA Program of China (Grants Nos. 2022SKA0110200 and 2022SKA0110203), the National Natural Science Foundation of China (Grants Nos. 12473001, 11975072, 11875102, and 11835009), the China Manned Space Project (Grant No. CMS-CSST-2021-B01), and the National 111 Project (Grant No. B16009).

This research has made use of data or software obtained from the Gravitational Wave Open Science Center (gwosc.org), a service of the LIGO Scientific Collaboration, the Virgo Collaboration, and KAGRA. This material is based upon work supported by NSF's LIGO Laboratory which is a major facility fully funded by the National Science Foundation, as well as the Science and Technology Facilities Council (STFC) of the United Kingdom, the Max-Planck-Society (MPS), and the State of Niedersachsen/Germany for support of the construction of Advanced LIGO and construction and operation of the GEO600 GW detector. Additional support for Advanced LIGO was provided by the Australian Research Council. Virgo is funded, through the European Gravitational Observatory (EGO), by the French Centre National de Recherche Scientifique (CNRS), the Italian Istituto Nazionale di Fisica Nucleare (INFN) and the Dutch Nikhef, with contributions by institutions from Belgium, Germany, Greece, Hungary, Ireland, Japan, Monaco, Poland, Portugal, Spain. KAGRA is supported by the Ministry of Education, Culture, Sports, Science and Technology (MEXT), Japan Society for the Promotion of Science (JSPS) in Japan; National Research Foundation (NRF) and Ministry of Science and ICT (MSIT) in Korea; Academia Sinica (AS) and National Science and Technology Council (NSTC) in Taiwan.

The data analyses and results visualization in this work made use of \texttt{BILBY} \cite{Romero-Shaw:2020owr, Ashton:2018jfp}, \texttt{dynesty} \cite{Speagle:2019ivv}, \texttt{LALSuite} \cite{LALSuite}, \texttt{Numpy} \cite{Harris:2020xlr, vanderWalt:2011bqk}, \texttt{Scipy} \cite{Virtanen:2019joe}, and \texttt{matplotlib} \cite{Hunter:2007ouj}.

\end{document}